\documentclass[twocolumn, aip, reprint]{revtex4-2}
\usepackage{lineno} 
\usepackage{appendix}
\usepackage{amsmath}
\usepackage{amssymb}
\usepackage{booktabs}
\usepackage{graphicx}
\usepackage{dcolumn}
\usepackage{bm}
\usepackage{xcolor}
\usepackage{array}
\usepackage{tabu}
\usepackage{hhline}
\usepackage{soul}
\usepackage{subfigure}
\usepackage{braket}
\usepackage{float}

\newcommand{\svtrb}{\langle \sigma_{\rm{tot}} \, v \rangle_{{\rm{Rb+}}X}}

\newcommand{\svtrbs}{\langle \sigma_{\rm{tot}} \, v \rangle_{^{87}{\rm{Rb+}}X}}

\newcommand{\svtrbi}{\langle \sigma_{\rm{tot}} \, v \rangle_{{\rm{Rb}}+i}}
\newcommand{\svtrbh}{\langle \sigma_{\rm{tot}} \, v \rangle_{\rm{Rb+H_2}}}

\newcommand{\svtlix}{\langle \sigma_{\rm{tot}} \, v \rangle_{^{6}{{\rm{Li+}}X}}}

\newcommand{\svtliv}{\langle \sigma_{\rm{tot}} \, v \rangle_{^{7}{{\rm{Li+}}X}}}

\newcommand{\rxs}{R_{6,87}}
\newcommand{\rss}{R_{7,87}}

\newcommand{\rcsx}{R^{C6}_{6,87}}
\newcommand{\rcss}{R^{C6}_{7,87}}

\newcommand{\svtli}{\langle \sigma_{\rm{tot}} \, v \rangle_{{\rm{Li+}}X}}
\newcommand{\svtlii}{\langle \sigma_{\rm{tot}} \, v \rangle_{{\rm{Li+}}i}}
\newcommand{\svtlih}{\langle \sigma_{\rm{tot}} \, v \rangle_{\rm{Li+H_2}}}

\newcommand{\svlli}{\langle \sigma_{\rm{loss}} \, v \rangle_{{\rm{Li+}}X}}
\newcommand{\svlrb}{\langle \sigma_{\rm{loss}} \, v \rangle_{{\rm{Rb+}}X}}

\newcommand{\grb}{\Gamma_{\rm{Rb}}}
\newcommand{\gli}{\Gamma_{\rm{Li}}}

\newcommand{\grbz}{\Gamma_{\rm{(Rb,0)}}}
\newcommand{\gliz}{\Gamma_{\rm{(Li,0)}}}

\newcommand{\rnistexp}{R_{7,87}^{\rm{BExp}}}
\newcommand{\rnistthy}{R_{7,87}^{\rm{KT}}}
\newcommand{\rubcexp}{R_{6,87}^{\rm{SExp}}}

\newcommand{\svloss}{\left<\sigma_{\rm{loss}}\, v\right>}
\newcommand{\svtot}{\left<\sigma_{\rm{tot}}\, v\right>}
\newcommand{\svtotsixzero}{\left<\sigma_{\rm{tot}}\, v\right>_{\rm{C_6}}}

\newcommand{\svtotcorr}{\left<\sigma_{\rm{tot}}\, v\right>_{\rm{corr}}}

\newcommand{\Ng}{n_{\rm{X}}}

\newcommand{\kb}{k_{\rm{B}}}
\newcommand{\vp}{v_{\rm{p}}}

\newcommand{\bea}{\begin{eqnarray}}
\newcommand{\eea}{\end{eqnarray}}

\newcommand{\frecap}{f_\mathrm{recap}}

\newcommand{\dni}{\Delta n_i}
\newcommand{\dnx}{\Delta n_X}

\newcommand{\ggli}{\dni \svtlii}
\newcommand{\gglix}{\dnx \svtli}

\newcommand{\ggrb}{\dni \svtrbi}
\newcommand{\ggrbx}{\dnx \svtrb}

\begin{document}

\title{Cross-calibration of quantum atomic sensors for pressure metrology} 


\author{Erik Frieling$^1$, Riley A. Stewart$^1$, James L. Booth$^2$ and Kirk W.~Madison$^1$.}
\affiliation{Department of Physics and Astronomy, University of British Columbia, 6224 Agricultural Road, Vancouver, B.C., V6T 1Z1, Canada} 
\affiliation{Department of Physics, British Columbia Institute of Technology, 3700 Willingdon Avenue, Burnaby, B.C. V5G 3H2, Canada}

\date{\today}

\begin{abstract}
{

Quantum atomic sensors have shown great promise for vacuum metrology. Specifically, the density of gas particles in vacuum can be determined by measuring the collision rate between the particles and an ensemble of sensor atoms.  This requires preparing the sensor atoms in a particular quantum state, observing the rate of changes of that state, and using the total collision rate coefficient for state-changing collisions to convert the rate into a corresponding density.   The total collision rate coefficient can be known by various methods including by quantum scattering calculations using a computed interaction potential for the collision pair, by measurements of the post-collision sensor-atom momentum recoil distribution, or empirically by measurements of the collision rate at a known density.  Observed discrepancies between the results of these methods calls into question their accuracy.  To investigate this, we study the ratio of collision rate measurements of co-located sensor atoms, $^{87}$Rb and $^6$Li, exposed to natural abundance versions of H$_2$, He, N$_2$, Ne, Ar, Kr, and Xe gases. This method does not require knowledge of the test gas density and is therefore free of the systematic errors inherent in efforts to introduce the test gas at a known density. Our results are systematically different at the level of 3 to 4\% from recent theoretical and experiment measurements.  This work demonstrates a model-free method for transferring the primacy of one atomic standard to another sensor atom and highlights the utility of sensor-atom cross-calibration experiments to check the validity of direct measurements and theoretical predictions.
}
\end{abstract}


\maketitle


\section{Introduction}
Ensembles of cold atoms have been proposed and investigated as a drift-free sensor of absolute pressure and particle flux measurements in vacuum and at ambient temperatures.\cite{booth2011,madison2012,arpornthip2012,yuan2013,moore2015,makhalov2016,makhalov2017,scherschligt2017,scherschligt2018,xiang2018,booth2019,shen2020,shen2021,barker2021,zhang2022,ehinger2022}  
The number density of the 
gas species, $X$, at temperature $T$, $n_X$, can be determined by measuring the collision rate between the sensor atoms and the ambient atoms and molecules via,
\bea
\Gamma(T) = n_X \svtot(T).
\label{eq:mastergamma}
\eea
The total collision rate coefficient, $\svtot(T)$, is the product of the relative collision speed, $v$, and the speed-dependent total collision cross-section, $\sigma_{\rm{tot}}(v)$.  For collisions between gas particles at room-temperature, $T \simeq 300$~K, and a laser-cooled sensor atom ensemble, the relative speeds are well-approximated by the Maxwell-Boltzmann (MB) velocity distribution of the gas \cite{HeatingPaper}. The brackets, $\langle \, \rangle$, indicate an average of the product over the MB velocity distribution at ambient temperature, $T$. In order to determine the number density at temperature $T$, the total collision rate coefficient, $\svtot(T)$, must be known.

The 
total collision rate coefficients can be obtained from quantum scattering calculations (QSC).\cite{makrides2019,makrides2020,makrides2022,makrides2022a,klos2023}  The quantum scattering computations used to determine the collision cross-sections rely on electronic potential energy surfaces (PESs), which must be determined by \textit{ab initio} quantum chemistry computations. However, sufficiently accurate PESs may be limited to molecular species with a small number of active degrees of freedom.
Alternatively, empirical estimates of the total collision rate coefficients, $\svtot(T)$, can be obtained from measurements with a gas of known density.\cite{barker2023,Eckel2023} This approach is limited to gas species compatible with the operation of existing orifice flow pressure standards.

A third method is to use the sensor atom collision recoil energy (or momentum) distribution to determine the total collision rate coefficient using the quantum diffractive collision universality law.\cite{booth2019,madison2018,shen2020,shen2021}  
The validity of this latter approach was previously shown for heavy collision partners including Rb+N$_2$ and Rb+Rb collisions \cite{stewart2022}, but deviations from the universality law were found for low mass test species.\cite{shen2023}  
The results of these three methods have been found to agree at the level of a few percent with the exception of a few special cases.\cite{barker2023,Eckel2023}  
Understanding the reason for the observed disagreements is essential.

The main goal of this work is to report the results of a fourth method for validating prior theoretical and experimental work. 
We perform direct experimental comparisons of the  total collision rate coefficients of co-located cold ensembles of $^{87}$Rb and $^6$Li atoms exposed to the test gas species, H$_2$, He, Ne, N$_2$, Ar, Kr and Xe.  We also present a comparison of the total collision rate coefficients when the sensor atoms are exposed to a sample of parahydrogen. This same method was used to compare $\svtrbh$ and $\svtlih$ in a previous study where the partial pressure of H$_2$ was varied by heating a non-evaporable getter (NEG)\cite{shen2023}. This allowed us to determine the ratio of the total collision rate coefficients~:~$\svtlih/\svtrbh = 0.83(5)$ for an ambient temperature of 300(2) K. The uncertainty in the ratio is the statistical uncertainty (type A) derived from fitting the loss rate data.
 

This method of comparing the total collision rate coefficients is model-independent and is free of some of the systematic errors inherent with other methods.  
In particular, the fact that the two sensor atom traps are co-located ensures they are both exposed to the same baseline environment and to the same test gas density. These conditions cannot be guaranteed for measurements comparing a cold atom vacuum standard to a calibrated ion gauge \cite{shen2020,shen2021} or to a flowmeter and dynamic expansion system where co-location is not possible.\cite{barker2023,Eckel2023}
In addition, this co-location measurement provides a systematic-error free method for transferring the primacy of one atomic standard to another sensor atom.

The ratio of the  total collision rate coefficients for various test gases, $X$, is,
\begin{equation}
R_X(T) = \frac{\svtli(T)}{\svtrb(T)}.
\label{eq:RXdefn}
\end{equation}
Our values are generally consistent with prior experimental and theoretical work.\cite{barker2023, Eckel2023, klos2023}
However, we do find statistically significant disagreements for certain test gas species.  In prior work, Barker \emph{et al.}~report collision-induced loss rates of $^7$Li and $^{87}$Rb from magnetic traps exposed to test gases whose density was set by a vacuum standard based on a flow meter and a dynamic expansion system. \cite{barker2023,Eckel2023}  Our findings are consistent with these prior measurements for He, but differ for the heavier species tested (Ne, N$_2$, Ar, Kr, and Xe). K{\l}os and Tiesinga report  total collision rate coefficients found from quantum scattering calculations. \cite{klos2023}
Our findings are consistent with these calculations for two of the light species tested, H$_2$ and Ne, but appear to indicate that the calculations may have systematically underestimated $R_X$ by approximately 3.4\% for He, N$_2$, Ar, Kr, and Xe.  We believe this work indicates the need for further refinement of the different methods of determining $\svtot(T)$.

\section{Experimental Procedure}\label{exp}

\begin{figure}
    \centering
    \includegraphics[width=\linewidth]{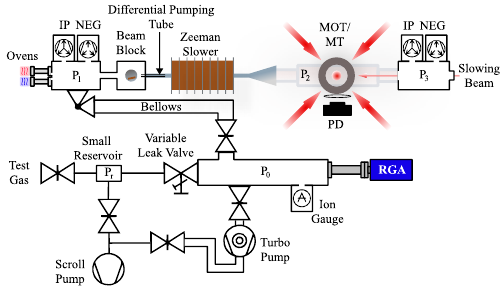}
    \caption{
        Schematic of the experimental apparatus. 
        The test gas is introduced into the vacuum system through a variable leak valve, resulting in a pressure $P_0$ in the first chamber set by the leak rate of the valve and the pumping rate of the turbo molecular pump.  
        The test gas is introduced from a small reservoir filled to a pressure $P_r\sim10^5$~Pa from a gas bottle. 
        The reservoir is pumped down using the scroll pump and filled for 2-3 cycles to purge any previous gases and ensure the selected test gas purity.
        A residual gas analyzer (RGA) is used to characterize the background gas constituents of $P_0$ and to verify that the leaked gas is free of contaminants.
        The gas flows through a 1.5 m long bellows into the oven chamber, which contains the Li and Rb ovens, a non-evaporable getter (NEG) and an ion pump (IP), reaching a pressure ${P_1}$.
        The atomic beams coming from the neighboring Li ($T_\mathrm{oven}=400\;^{\circ}$C) and Rb ($T_\mathrm{oven}=100\;^{\circ}$C) ovens are decelerated using a Zeeman slower and captured in a magneto-optic trap (MOT). The atoms are subsequently transferred to a magnetic trap (MT) in which they suffer collision-induced losses from the test gas, now at $P_2$ due to differential pumping.
        The beam block can be rotated to block the directed atomic flux from the ovens during the measurement. 
        It is located next to a cold finger kept at $5^\circ$C that collects some of the excess Rb and Li atoms to avoid contamination of the vacuum system. 
        The atoms remaining in the MT after exposure are recaptured in the MOT, and their fluorescence is measured using a photodiode (PD). 
        }
    \label{fig:apparatus}
\end{figure}

\begin{figure*}[t]
    \includegraphics[width=\textwidth]{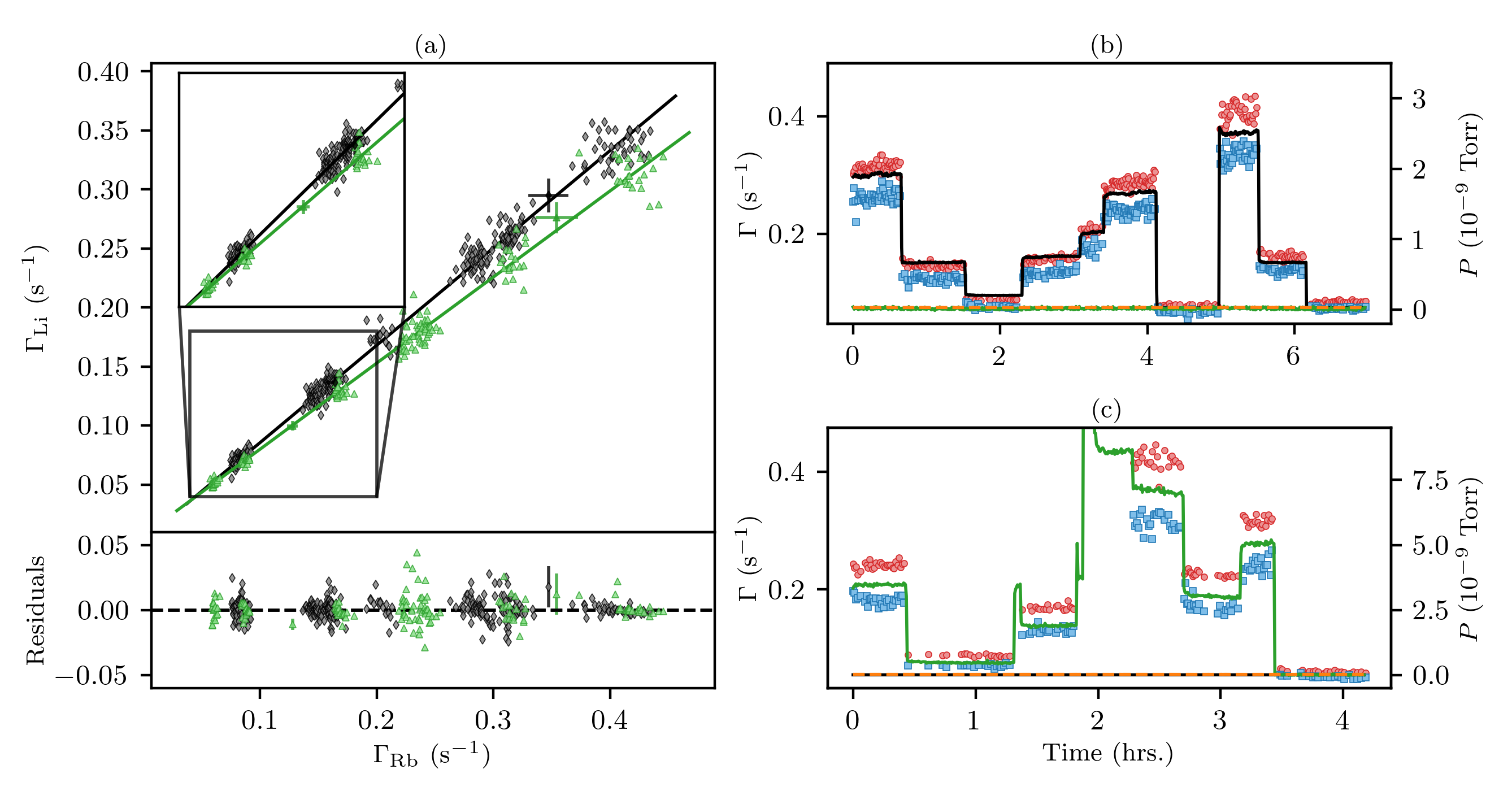}
\caption{
(a) Magnetic trap loss rates for $^6$Li and $^{87}$Rb atoms, plotted as the ordered pair $(\grb,\gli)$, when exposed to H$_2$ (black diamonds) and He (green triangles) gas. The measurements are fit to lines (same color as markers) whose slope provides the ratio of the loss rates and thus the ratio of the total collision rate coefficients. We find $R_X =0.824(5) \; (0.729(6))$ for H$_2$ (He). These $R$ values are derived from the raw measurements  $R^{\rm{meas}}_X$ = 0.825(5) (0.729(6)) corrected for finite trap depth effects as explained in appendix section \ref{app:energy}.
Example error bars are shown for a few samples with the rest suppressed for readability.
Results for all test gases are summarized in table \ref{tab:results}.
(b) and (c) show the loss rate measurements for Li (blue squares) and Rb (red circles) from the shallow magnetic trap together with the inferred partial pressure of H$_2$ (black), He (green) and N$_2$ (orange dashed) in the MOT/MT chamber as measured by a residual gas analyzer (RGA) and scaled using a calibration factor (see Appendix \ref{app:gases}).}
\label{fig:gvgplot}
\end{figure*}


In this work, the ratio $R_X(T)$ is measured directly by confining Rb and Li atoms to the same spatial region. Prior to introducing any test gas, $X$, to the vacuum system, the sensor atom ensembles exhibit baseline trap loss rates.  These baseline loss rates are due to collisions with common residual background gases, such as H$_2$, and intrinsic trap loss mechanisms.

We then leak into the chamber the test gas $X$. This results in a fixed (but unknown) density, $n_X$, at ambient temperature, $T = 298(2)$ K, in the vacuum system resulting in trap loss rates of
\bea
\grb\left(n_X, T\right) & = & n_X \svtrb(T) + \grbz
\label{eq:theRbgamma}
\eea
\bea
\gli\left(n_X , T\right) & = & n_X \svtli(T) + \gliz.
\label{eq:theLigamma}
\eea
where $\grbz$ and $\gliz$ are the baseline loss rates excluding contributions from collisions with particles of type $X$. These expressions describe the trap loss rate of sensor atoms confined in sufficiently shallow traps such that every collision with species, $X$, liberates a trapped atom.

The loss rates, expressed in Eq.~\ref{eq:theRbgamma} and \ref{eq:theLigamma}, are measured, pair-wise, for a range of test gas densities at temperature, $T = 298(2)$ K. For all values of $n_X$, the loss rates are linearly related by the value of $R_X$ as
\bea
\gli\left( n_X, T\right) &=& R_X(T)\cdot \grb\left(n_X, T\right) + b.
\label{eq:glivsgrb}
\eea
The result is that a plot of the measured loss rate pairs, $\gli$ versus $\grb$, for all values of the test gas density, yields a linear relationship with slope, $R_X(T)$, and a constant intercept, $b = \left[\gliz - R_X(T)\grbz \right]$. 

The advantage of this method of determining $R_X$ is that an exact knowledge of the test gas density or residual background composition and the background density are not required.  Instead, it is sufficient that these densities remain constant for each measurement pair. This technique is in contrast with prior experimental work that relied on another pressure standard to set a known test gas density \cite{barker2023,Eckel2023}.


The apparatus used to perform these loss rate measurements of both Rb and Li atoms 
is very similar to one described in previous work, although the current iteration is a complete rebuild.\cite{bowden2016,shen2023}
In the rebuilt version, an additional vacuum chamber with a variable leak valve, ion gauge and residual gas analyzer (RGA) was connected near the oven section of our apparatus. This allows us to control the partial pressure $P_X = n_X \kb T$ of the test gas $X$ introduced through the leak valve, where $\kb$ is Boltzmann's constant

The temperature of the gases inside the vacuum was inferred from a measurement of the room temperature by a single digital thermometer above the apparatus.  We assume the test gas is at thermal equilibrium with the room.  As discussed in the Appendix, the common temperature variation of the total collision rate coefficients renders their ratio insensitive to temperature variations.  Specifically, even an uncertainty of $\pm 10$ K on an ambient temperature of 300 K results in an uncertainty of $R_X$ of less than 0.06\%. Thus, there is no requirement to measure the temperature with high accuracy or precision.

A schematic of the apparatus is shown in Fig.~\ref{fig:apparatus}.
When introducing a new gas, we start by evacuating a small reservoir connected to the variable leak valve and the scroll pump.
Subsequently, we purge the reservoir by filling the reservoir with the test gas at a pressure of approximately $10^5$ Pa and evacuating it again for 2-3 purge cycles.
The reservoir is then filled to a pressure $P_r \sim 10^5$~Pa and sealed.
We then vary the partial pressure of the test gas in the magnetic trap chamber, in the range from $10^{-8}$ to $10^{-7}$~Pa, by changing the small leak rate using the variable leak valve.
The RGA is used to verify the leaked gas is free of contaminants.

The oven section and magnetic trap section, where the trap loss measurements are performed, are connected by a 12~mm long, 6~mm diameter differential pumping tube with an estimated Hydrogen conductance of 0.85L/s\cite{bowden2014}.
We turn off the ion pump in the oven section (where the pressure is $P_1$ as shown in Fig.~\ref{fig:apparatus}) to prevent sputtering and pressure fluctuations due to Argon instability when pumping heavy noble gases.\cite{jousten2016}
The much lower pressure in the magnetic trap section eliminates the need to do the same for the ion pump located there.  Instead, we leave this second ion pump on, because it lowers the baseline pressure achieved when not flowing the test gas and, as a result, improves the lower bound on the measured loss rates and enhances the experimental precision achieved in $R_X$.  Because the statistical estimate for the linear slope, $R_X$, is optimized by making measurements that are spaced far apart in the independent variable (here the Rb loss rate), we lower the baseline pressure to expand the span of the loss rate measurements, which are bounded from above at a pressure where the loss is so high that the signal-to-noise becomes severely compromised.  In addition, we verified that turning off this ion pump does not alter the measured value of $R_X$.

For both species, co-located magneto-optic traps (MOTs) capture atoms from thermal, parallel atomic beams that are decelerated and cooled by a Zeeman slower.\cite{bowden2016} To ensure a similar number of atoms are loaded for each experimental shot, the Zeeman slowing light is extinguished when the fluorescence from the MOT reaches a set value $V_\mathrm{MOT}$.  The atoms are then optically pumped to the desired hyperfine state and all light is extinguished, at which point they are confined by the quadrupole magnetic field.  We then ramp the magnetic field gradient in 20~ms to the hold gradient used for the MT phase. After a variable hold time in the MT, the atoms are recaptured in the MOT and their fluorescence $V_\mathrm{MT}(t)$ is recorded using a photodiode. To improve the signal to noise, we determine the recaptured fraction $\frecap(t)=V_\mathrm{MT}(t)/V_\mathrm{MOT}$ as described previously. \cite{shen2023}
This quantity is insensitive to the residual variations in the number of atoms loaded into the MOT.

Glancing collisions which do not induce trap loss can lead to heating of the ensemble and, consequently, influence the measured loss rate in a non-trivial way (see appendix \ref{app:energy}). To minimize this effect,
the Li and Rb atoms are confined in a very shallow magnetic trap for these measurements.  The Rb atoms are confined with an axial magnetic field gradient of $b'=35$ G/cm, and the trap depth is limited to $U_\mathrm{Rb}/k_B=125~\mathrm{\mu K}$ by applying radio-frequency (RF) radiation repeatedly sweeping between 20 MHz and 40 MHz for 2 s at the end of the variable hold time.  We verified that sweeping the RF from 0 to 40 MHz completely empties the trap.  The RF induces spatially localized transitions between the trapped, weak-field seeking $\ket{F=1,m_F=-1}$ and the untrapped $\ket{F=1, m_F=0}$ and anti-trapped, strong-field seeking $\ket{F=1, m_F=1}$ magnetic Zeeman states.  More details are provided in the appendix \ref{app:energy}. For $^{6}$Li, the atoms are confined in an axial gradient $b'=100$ G/cm.
We do not apply any RF for $^{6}$Li, since the field dependence of the magnetic moment of the $\ket{F=1/2,m_F=-1/2}$ state limits the trap depth to $U_\mathrm{max}/k_B=314~\mathrm{\mu K}$.\cite{shen2023}

Because the optimal Zeeman+MOT loading and MT trapping parameters are so different for the two sensor atoms, we alternate between loss rate measurements instead of making a simultaneous measurements of a MT loaded with both sensor atoms.  Since we require that $\Ng$ be the same for each sensor atom loss rate measurement, we perform measurements at a series of different but constant leak rates.  To mitigate the effect of any slow drift of the background gas density, we minimize the time between sensor atom measurements by performing two-point measurements, alternating between trapped species.  We measure the recapture fraction at $t=0$ and at $t=t_1$, and we compute the loss rate as $\Gamma=\ln\left[\frecap(t=t_1)/\frecap(t=0)\right]/t_1$.  We interlace the measurements for the two species such that the two-point lifetime measurement for the two species is overlapped in time.  Thus a single $\left(\grb, \gli\right)$ point is collected as follows:
\begin{enumerate}
    \item Measure $\frecap(t=0)$ for Li
    \item Measure $\frecap(t=0)$ for Rb
    \item Measure $\frecap(t=t_1)$ for Li
    \item Measure $\frecap(t=t_1)$ for Rb
\end{enumerate}
To optimize the sampling, we choose a hold time for each species such that $t_1 \equiv1.5/\bar{\Gamma}$, where $\bar{\Gamma}$ is an estimate of the Rb or Li loss rate that is updated based on the average of the previous five measurements.  In addition, the order in which the $t=0$ and $t=t_1$ measurements are taken is randomized to mitigate the effect of systematic drifts of the MOT to MT transfer efficiency.  We also observed that the decay of the recapture fraction, even at the very lowest background pressures where non-linearities in the decay would be most apparent, is purely exponential (See Appendix section \ref{app:nonlinear}).  The consequence is that a two-point measurement is sufficient to determine the total loss rate at any test gas pressure.

Finally, we note that for this study, we use $^6$Li instead of $^7$Li sensor atoms.  Because the $^6$Li atom is a composite Fermion, when we prepare and magnetically trap a spin polarized sample in the $|F=1/2, m_F=-1/2\rangle$ state at a temperature below 300 $\mu$K, intra-trap collisions do not occur.  This is because, s-wave collisions are forbidden due to the Pauli principle and p-wave collisions are frozen out at this temperature.  Because collisions between sensor atoms can lead evaporation losses that confound the determination of the losses due to collisions with background particles, such collisions are problematic for precision vacuum metrology.  Efforts to minimize sensor atom collisions have been described previously. \cite{shen2021,stewart2022,shen2023,HeatingPaper}  Here, these efforts are unnecessary because the $^6$Li ensemble is a true experimental realization of the idealized model of a non-interacting trapped sensor atom ensemble.

\section{Discussion of Results}
In Fig.~\ref{fig:gvgplot} we plot the ordered pairs $(\grb,\gli)$ for different values of the H$_2$ and He densities in a 2D scatter plot where the abscissa is the Rb loss rate and the ordinate is the Li loss rate.  The measurements for H$_2$ and He were chosen as examples, because they represent the highest and lowest values of $R_X$ respectively.
We fit these points to a linear model using orthogonal distance regression (ODR) and extract the slope $R$.
The individual points are weighted by the statistical uncertainty, which we estimate by characterizing the variation in $\frecap$ at constant pressure and using uncertainty propagation to find the uncertainty in $\Gamma$.
In Fig.~\ref{fig:gvgplot} these uncertainties are suppressed for visual clarity, except for a few selected points.

Figs.~\ref{fig:gvgplot}b and \ref{fig:gvgplot}c show the same data as a time series, together with the partial pressures measured by the RGA. 
For each gas under test, the partial pressure was increased and decreased several times in random order during each experimental run to minimize the effects of any systematic, slow variation in the background pressure on the measurement. 
We also repeated each measurement on three different days and observed the same values for $R_X$ to within 1\%.

\addtolength{\tabcolsep}{+5pt}    
\begin{table*}
\caption{\label{tab:results} The measured total collision rate coefficient ratios, $R^{\rm{meas}}_{6,87} = \svlli/\svlrb$ , and the trap-depth-corrected ratios $\rxs~=~\svtlix/\svtrbs$ in this work. $R^{\rm{KT}}_{7,87}$  are calculated from quantum scattering calculations\cite{klos2023} of $\svtliv$ and $\svtrbs$. $R^{\rm{BExp}}_{7,87}$ are computed from previous measurements\cite{Eckel2023} of $\svtliv$ and $\svtrbs$  . $R^{\rm{SExp}}_{6,87}$ is the value from our prior direct cross-species measurement \cite{shen2023} and $R^{\rm{th.}}_{6,87}$ is the quantum scattering computation reported in the same paper. The final column shows the values for $R^{\rm{C6}}_{6,87}$ predicted by Eq.~\ref{eq:svtot60}.
Agreement with prior theory, $\rnistthy$, is found for $X$=H$_2$ and Ne; however, statistically significant deviations both from previous theory and experiment are found for other species. All of the $R$ values listed here have been adjusted for a room temperature of 298(2)~K, as described in the text.
These results are also plotted in Fig.~\ref{fig:Rcomparisonplot}. The statistical uncertainties (type A) are listed in brackets beside each entry. The systematic uncertainties (type B), as discussed in Appendix A, are much lower and their maximum values are listed in Table~\ref{Tab:EtypeB}. 
}
\begin{tabular}{|l|cc|cc|cc|c|}
    \hline
     & \multicolumn{2}{c|}{(This work)} & & & & & \\
    Species &   $R^{\rm{meas}}_{6,87}$  &  $\rxs$ & $\rnistthy$ (\cite{klos2023}) & $\rnistexp$ (\cite{barker2023,Eckel2023}) & $\rubcexp$ (\cite{shen2023}) & $\rxs^{\mathrm{(th.)}}$ (\cite{shen2023}) & $\rcsx$\\
    \hline
      $p $H$_2$ & 0.828(6) &  0.827(6) &                0.82(3)\phantom{2} &                            &                &                   0.869 & 0.801(3) \\
    
      H$_2$ & 0.825(5) &  0.824(5)  &               0.82(3)\phantom{2} &                            &                  0.83(5) &                   0.869 & 0.801(3)\\
         He & 0.729(6) &  0.729(6)    &          \  0.69(2) \phantom{2} &                    0.73(2)\phantom{2} &                          &                   & 0.775(1)      \\
         Ne & 0.798(5) &  0.797(5)    &             0.78(10)  &                  \ 0.74(2) \phantom{2} &                          &                      & 0.778(2)   \\
      N$_2$ & 0.798(5) &  0.796(4)   &            \ 0.764(14) &                    0.75(3)\phantom{2} &                          &                       & 0.789(1)  \\
         Ar & 0.798(5) &  0.796(5)   &             0.768(4) &                  \ 0.721(13) \phantom{2} &                          &                       & 0.788(1)  \\
         Kr & 0.798(5) &   0.795(5)   &            0.769(4) &                 \  0.78(2) \phantom{2} &                          &                        & 0.797(2) \\
         Xe & 0.803(4) &  0.799(4)    &            0.779(8) &                    0.77(2)\phantom{2} &                          &                      & 0.803(2)   \\
    \hline
    \end{tabular}

\end{table*}
\addtolength{\tabcolsep}{-5pt}    

To compare the current results with previous work, we must compensate the measured loss rates for the finite trap depths for each sensor atom ensemble, $U_{\rm{Li}}$ and $U_{\rm{Rb}}$,  at which the measurements were carried out. Namely, the quantity that we measure,
\bea 
R^{\rm{meas}}_X\left(T, U_{\rm{Li}}, U_{\rm{Rb}}\right) &=& \frac{\svlli(T,U_{\rm{Li}})}{\svlrb(T,U_{\rm{Rb}})}
\label{eq:Rmeas}
\eea
becomes $R_X(T)$ in the limit that the $U_{\rm{Li}} \rightarrow 0$ and $U_{\rm{Rb}} \rightarrow 0$.  K{\l}os and Tiesinga (KT)\cite{klos2023} report a convenient quadratic description for extrapolating $\svtot$  from the measured $\svloss(T,U)$ (labelled $K(T)$ and $L(T,U)$, respectively, in their work) for shallow traps,
\bea 
\svloss(T,U) &=& \svtot(T) - a_{\rm{gl}}(T)\cdot U + b_{\rm{gl}}(T)\cdot U^2. \nonumber \\
\label{eq:KTsvloss}
\eea
Using the KT $a_{\rm{gl}}$ and $b_{\rm{gl}}$ coefficients, we determine the $R_X(T = 298\ K)$ extrapolated from the theoretical values quoted by K{\l}os and Tiesinga\cite{klos2023}. 
(See Appendix \ref{app:energy} and Table~\ref{tab:finiteEnergy}.) 
Since we use such shallow magnetic traps to confine the sensor atoms, the trap depth correction for the Li$+X$ measurements is less than 0.1\%, and the Rb$+X$ correction is less than 0.5\% for all collision partners. 
Similarly, trap depth corrections are applied to the Li$+X$ loss rate coefficients measured by Barker \textit{et al} (Bexp) \cite{barker2023, Eckel2023}.

The data presented here were collected for test gases at a temperature of 298(2) K. By contrast, the KT theoretical values were computed for a test gas temperature of 300 K and the Barker \textit{et al} (BExp) \cite{barker2023, Eckel2023} measured the Li$+X$ loss rate coefficients at 300.2(2.9) K and the Rb$+X$ loss rate coefficients at 295.2(3) K.  Therefore, the BExp results were temperature corrected to $T=298$ K following the prescription of K{\l}os and Tiesinga (KT)\cite{klos2023}. Namely,
\bea
\svtot(T) &=& K_0 + K_1\cdot\left(T - 300 \; \mathrm{K}\right)
\label{eq:KTKcorrect}
\eea
where $K_0$ is the loss rate coefficient at $T = 300$~K, and $K_1$ is the linear temperature correction coefficient. These corrections allowed us to compute the KT theortical and BExp measured ratios $R_X(298\ K)$ values that can be compared to our present data directly.

\begin{center}
\begin{figure}
   \centering
   \includegraphics[width=\columnwidth]{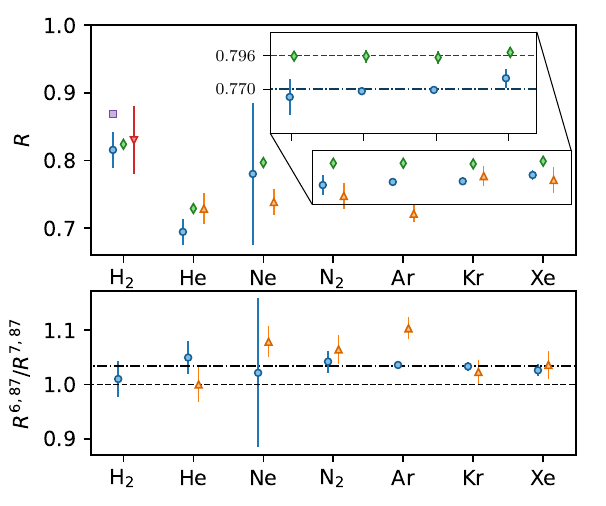} 

\caption{Top panel: $R_{6,87} = \frac{\gli}{\grb} = \frac{\svtli}{\svtrb}$ as measured in this work (green diamonds) for variety of background gases, compared to $R^{\rm{BExp}}_{7,87}$ (\cite{Eckel2023}) (orange triangles) and quantum scattering calculations values, $R^{\rm{KT}}_{7,87}$ (\cite{klos2023}) (blue circles).
Our previous experimental result (red nabla) and theoretical prediction (purple square) for H$_2$ are also included.\cite{shen2023} 
These data are summarized in Table~\ref{tab:results}. The inset shows the $R_{6,87}$ data and the $R^{\rm{KT}}_{7,87}$ values for N$_2$, Ar, Kr, and Xe. The dashed lines  are the average values, $\left<R\right>$ for these collision partners. Note that the $R_{6,87}$ and $R^{\rm{KT}}_{7,87}$ are remarkably constant for the heavier collision partners. Bottom panel: 
Ratio of $R_{6,87}$ determined using our cross-species calibration method (this work) to KT theory\cite{klos2023} (blue circles) and the ratio of $R_{6,87}$ to Bexp measurements \cite{Eckel2023} (orange triangles) for the different collision partners. The dashed line corresponds to a value of 1.00 (perfect agreement). The dot-dashed line is the average, $\left<R_{6,87}/R^{\rm{KT}}_{7,87}\right> = 1.034(7)$ for X = N$_2$, Ar, Kr, and Xe.
}
\label{fig:Rcomparisonplot}
\end{figure}
\end{center}

The results of our measurements are summarized in Tab.\ref{tab:results} and Fig.~\ref{fig:Rcomparisonplot}, where they are compared to recent theoretical and experimental results\cite{klos2023, barker2023, shen2023,Eckel2023}. We note that our measurements utilize $^6$Li and $^{87}$Rb atoms and thus we report $\rxs \equiv \frac{\svtlix}{\svtrbs}$, whereas prior theoretical and experimental results are for $^7$Li and $^{87}$Rb atoms and thus provide  $\rss \equiv \frac{\svtliv}{\svtrbs}$.

The data set is observed to separate into two groups: the higher mass collision partners for which the ratio is almost constant (with an average value $\rxs = 0.797(2)$ for Ne, N$_2$, Ar, Kr, and Xe), and the lower mass collision partners, H$_2$, He, for which the $R_{6,87}$ are distinct from this constant value.
  
The $\rxs$ values reported here for H$_2$ and Ne are consistent with the quantum scattering calculations (QSC) of K{\l}os and Tiesinga\cite{klos2023} (KT). 
Specifically, the measured value for H$_2$, $\rxs = 0.824(5)$, agrees with the KT QSC value, $\rnistthy = 0.82(3)$.  It is also in agreement with our previous measurement, $\rubcexp  = 0.83(5)$.  Our earlier experimental measurement also employed co-located sensor ensembles, but relied on a different source for the H$_2$ gas (the H$_2$ outgassing induced by heating a non-evaporable getter pump).  The measured value for He,  $\rxs = 0.729(6)$, agrees with the updated BExp value\cite{barker2023,Eckel2023}, $\rnistexp = 0.73(2)$, and is within 2 standard deviations of the KT QSC value, $\rnistthy = 0.69(2)$.  Finally, the value for Ne, $\rxs = 0.797(5)$ is in agreement with the KT QSC value, $\rnistthy = 0.78(11)$, but is outside 2 standard deviations of the BExp measurement of $\rnistexp = 0.74(2)$.  By contrast, the $\rxs$ values for the heavier species (N$_2$, Ar, Kr, and Xe) are systematically larger than either the KT QSC or the BExp experimental values (see Tab.~\ref{tab:results}).  We discuss these discrepancies and potential causes here and in the Appendix.

A unique advantage of our experimental technique and the data we have obtained is that the Li and Rb sensor atoms are exposed to exactly the same 
test gas pressure, since the traps are co-located.  In the experimental work of Barker \textit{et al}\cite{barker2023, Eckel2023}, $\svtot(T)$ was deduced from measurements of the loss rate at test gas densities, $n_X$, created by an orifice flow standard. We hypothesize that pressure gradients may have existed between the orifice flow standard output and the locations of the two separate CAVS sensor assemblies connected to the standard. Indeed, the ambient temperatures reported for these two sensor ensembles were different, indicating variations in the local test environments. Such gradients could result in a systematic error in the inferred pressure and thus the $\svtot(T)$ values reported.  As the lightest noble gas, He has the highest conductance through the vacuum system and does not exhibit significant out-gassing or surface adsorption, so pressure gradients are expected to be smallest for this species.  This may account for the close agreement between the experimental values. By contrast, we observe larger discrepancies between the $\rnistexp$ and the $\rxs$ values reported here for the heavier species (see Tab.~\ref{tab:results}), consistent with this hypothesis.

A remarkable feature is the almost identical $R_{6,87}$ values for the highest mass collision partners, N$_2$, Ar, Kr,
and Xe. This same trend is seen in the quantum scattering computed values, $R^{\rm{KT}}_{7, 87}$, albeit with a different average
value. We believe that this collision partner independence of $R_{6,87}$ is related to universal nature of the
collisions.\cite{booth2019, shen2020, shen2021, shen2023} For these heavy collision partners, both
the loss rate variation with trap depth and the total collision rate coefficients are expected to be independent of the interaction potential at short range. In this case, the $\svtot$ values are principally determined by the long-range van der
Waals interactions (characterized by the dominant term, $-C_6/r^6$, where r is the inter-species separation). For the
purposes of this discussion, we can write the total collision rate coefficient as,
\bea
\svtot &=& \svtotsixzero + \svtotcorr
\label{eq:svtotapprox}
\eea
where $\svtotsixzero$ is the prediction based on the Jeffreys-Born approximation,\cite{booth2019}
\bea
\svtotsixzero &=& 8.49464 \left(\frac{C_6}{\hbar \vp}\right)^{\frac{2}{5}} \vp + \nonumber \\
& & \ \ \ \ \ \ \ \ 7.19729 \left(\frac{\hbar}{\mu}\right) \left(\frac{C_6}{\hbar \vp}\right)^{\frac{1}{5}}.
\label{eq:svtot60}
\eea
Here $\vp$ is the peak velocity of the Maxwell-Boltzmann distribution of the test gas species, and $\mu$ is the reduced mass of the colliding partners. The $\svtotcorr$ are the corrections introduced by the other long-range van der Waals interactions and the details of the core repulsion.  We note that because these corrections affect the quantum-mechanical scattering amplitude, the value of $\svtotcorr$ can be zero, positive or negative, corresponding to a constructive or destructive interference to the total cross section. Using this, we can write,
\bea
R &=& R^{\rm{C6}}\cdot \left[ \frac{ \left(1 + \alpha_{\rm{Li}+X} \right) }{\left(1 + \alpha_{\rm{Rb}+X} \right) } \right]
\label{eq:RC6terms}
\eea
where
\bea
R^{\rm{C6}} &=& \frac{ \svtotsixzero^{\rm{Li}+X}}{\svtotsixzero^{\rm{Rb}+X}}
\label{eq:RC6prediction}
\eea
and
\bea
\alpha &=& \frac{\svtotcorr^{\rm{A+X}}}{\svtotsixzero^{\rm{A+X}}}.
\label{eq:Tcorr}
\eea
Here A is the sensor atom, either Li or Rb.

The final column in Table~\ref{tab:results} shows the $\rcsx$ predictions using the $C_6$ values taken from Jiang \textit{et al}.\cite{jiang2015} Computing the average of our measured values for the heaviest species (N$_2$, Ar, Kr, and Xe) yields
\bea
\left< \rxs \right>&=& 0.7965(9),
\label{eq:RUBCheavy}
\eea
and appears to agree with the corresponding average value for the $\rcsx$ predictions,
\bea 
\left<\rcsx\right> &=& 0.795(5).
\eea
This result implies that the ratio of the correction terms in Eq.~\ref{eq:RC6terms} cancel and that $R^{\rm{C6}}$ is a good approximation for any two alkali sensor atoms. Thus, one can use the predicted $R^{\rm{C6}}_{\rm{A,B}}$ value and a measurement of $\svtot$ for one sensor atom collision-partner pair (A-X) to deduce the $\svtot$ corresponding to a different sensor atom (B-X).  We note that for $^7$Li:$^{87}$Rb, the average is $\left< \rcss \right>~=~0.792(5)$ owing to the $\mu$ dependence of the second term of Eq.~\ref{eq:svtot60}.

We also had the opportunity to measure $\rxs$ for $>$99.9\% pure para-hydrogen ($p$H$_2$) created using a closed cycle He fridge and a (FeOH)O magnetic converter at 14~K. \cite{10.1063/1.3072881}.  This gas was produced and provided to us by the group of Professor Takamasa Momose of the University of British Columbia.
We found no statistical difference in the value of $R$ for para-hydrogen compared to normal hydrogen, as shown in Table \ref{tab:results}.
Normal hydrogen is a mixture of ortho- and para-hydrogen with nuclear spins, I = 1 and 0, respectively. In the $\rm{X\ ^1\Sigma^+_{g}}$ ground molecular state, the wavefunction symmetry demands that I + J is even, where J is the rotational quantum number of the rovibronic state. Thus, ortho-hydrogen populates only the odd J rotational levels and para-hydrogen populates only the even rotational levels.
For glancing collisions there is insufficient anisotropy in the three body PESs to result in a noticeable change in rotational state required for an para-H$_2$ to ortho-H$_2$ conversion (or vice-versa). 
Thus we expected the $R_{6,87}$ (and $\svtli, \svtrb$) to be the same for normal H$_2$ and para-H$_2$, as our measurements confirm.

In the appendices, we discuss the systematic errors (type B errors) associated with the $\rxs$ values reported here. These are summarized in Table~\ref{Tab:EtypeB}. The largest systematic error is associated with the corrections due to the finite trap depth confining the sensor atoms deviating the measured value of $\Gamma_{\rm{loss}} = n\svloss$ from the target $\Gamma = n \svtot$. This systematic error has been removed from our quoted values of $R_{6,87}$ in Tab.~\ref{tab:results}, and it is well below the 1\% level for all collision partners for the trap depths employed in this study. Thus, we are confident that the discrepancies we observe indicate the need for the refinement of both theoretical and experimental determinations of $\svtot$.
\begin{table}[h]
\caption{Summary of the systematic error estimates (type B errors) associated with this work. The largest systematic errors arise from the finite trap depth of the confined sensor atoms shifting the $\Gamma^{\rm{meas}}_{\rm{loss}}$ away from $\Gamma_{\rm{tot}}$.  This uncertainty has been removed from the $R_{6,87}$ values quoted in Table~\ref{tab:results}. Detailed discussions of each uncertainty can be found in the Appendices.}
\begin{tabular}{lr}
\hline
Error Source & Estimate, $\frac{\delta R}{R}$ \vspace{3pt} \\ 
\hline 
\hline
\vspace{3pt}
 $\Gamma^{\rm{meas}}_{\rm{Loss}}$ deviation from $\Gamma_{\rm{tot}}$       & $<0.5$\% 
 \vspace{3pt} \\
 \hline
Ensemble Heating Induced Error      & $<0.004$\%     \\  
2-Body Intra-Trap Loss Uncertainty & $< 0.05$\% \\
Ambient Temperature Uncertainty& $< 0.012$\% \\
Baseline Gas Composition Variation & $<0.14$\% \\
\hline 
\hline
\end{tabular}
\label{Tab:EtypeB}
\end{table}

In Appendix B, we hypothesize that the systematic 3.4(7)\% difference between the measured $R_{6,87}$ and computed $R^{\rm{KT}}_{7,87}$ values for $X$ = N$_2$, Ar, Kr, and Xe is due to the incomplete suppression of the effects of glory oscillations via the Maxwell-Boltzmann (MB) averaging of $\svtot$ for Li-$X$ collisions. By contrast, the glory scattering is almost completely erased by the MB averaging for Rb-$X$ collisions. We support this by illustrating the very different glory oscillation behaviour between Li-Xe and Rb-Xe, by comparing the computed $\svtot_{\rm{KT}}$ values from the potentials reported by KT to a semi-classical (SC) interaction potential consisting solely of long range van der Waals terms, $\svtot_{\rm{SC}}$. The ratio of these total collision loss rate coefficients, $\svtot_{\rm{SC}}:\svtot_{\rm{KT}}$ is 1.0 for Rb-$X$ ($X$ = N$_2$, Ar, Kr, and Xe) collision partners, indicating a near complete erasure of glory oscillation effects. The corresponding ratio for is $\simeq 1.02$ for Li-$X$ collisions indicating some information about the glory scattering persists in these $\svtot$. Thus, the Li-$X$ $\svtot$ values are conveying some information about the details of the core portion of the interaction potentials. The 3.4(7)\% discrepancy could be a manifestation of some small inaccuracy of the potential used by KT to compute the Li-$X$ $\svtot_{\rm{KT}}$ values. 


The hypothesis that the potentials used to compute the Li$+X$ cross sections may need refinement is supported both by our experimental results and by the experimental values of $\svtot$ for $^{7}$Li collisions with the heavier collision partners reported by Barker \textit{et al.}\cite{barker2023}. The majority of the experimental values reported by Barker \textit{et al.} are systematically 2-4\% higher than the theoretical predictions.  The results presented here, with their attendant high precision, offer an exciting opportunity to examine the shape of the core repulsion for Li collisions experimentally.

\section{Conclusions}
The collision-induced loss rate of cold, trapped atoms from a shallow trap due to a test gas, $X$, can be written as, $\Gamma(T) = n_{\rm{X}} \svtot(T)$, where $n_{X}$ is the density of the test gas, and  $\svtot(T)$ is the total collision rate coefficient characterizing the collision interaction. A knowledge of the total collision rate coefficient allows the measured loss rate to be used to measure the gas density directly. Here we have measured the loss rates of $^{87}$Rb atoms and of $^6$Li 
atoms from shallow magnetic traps
when exposed to natural abundance versions of H$_2$, N$_2$, He, Ne, Ar, Kr, and Xe gases at T = 298(2) K. The loss rates were recorded in pairs, $\left( \Gamma_{\rm{Rb}+X}, \Gamma_{\rm{Li}+X} \right)$, for a series of gas densities, $n_{\rm{X}}$, for each gas species, $X$.  These loss rates, measured at trap depths $U_{\rm{Li}} < 0.314$ mK and $U_{\rm{Rb}}~<~0.125$~mK, were used to compute the collision rate coefficient ratio, $R_X(T) = \left<\sigma_{\rm{tot}} v\right>_{\rm{Li}+X}/\left<\sigma_{\rm{tot}} v\right>_{\rm{Rb}+X}$. The innovation and advantages of this method are (i) the cold trapped ensembles are co-located in the vacuum system ensuring that the two trapped ensembles experience the same static background gas environment and the same test gas densities free from any systematic errors that may be encountered for ensembles probing the vacuum at different locations, and (ii) the ratiometric method means that the density of the test gas need not be measured or known to obtain $R_X(T)$.  The only requirement is that the test gas density remain constant over the measurement duration (tens of seconds) for each density setting. This method provides an experimental technique for transferring the known or calibrated  loss rate coefficient for one species pair (A-X) to a second species pair (B-X), where $X$ is the test gas and A and B are two different trapped sensor atom ensembles.  This experimentally determined loss rate ratio, $R_X(T)$, also provides systematic-error free measurements that can be compared directly to \textit{ab initio} computations of the loss rate coefficients $\left<\sigma_{\rm{tot}}v \right>(T)$. We believe these measurements can be used to guide refinements of the theoretical potential energy surfaces describing the collision interactions. 

The measurements indicate that the loss rate ratios are constant for the heaviest species tested ($X$ = N$_2$, Ar, Kr, and Xe) with an average of $\left< R_{6,87} \right> = 0.7965(9)$.  This is remarkably close to the purely C$_6$ prediction $\left< R^{\rm{C6}}_{6, 87} \right> = 0.795(5)$ for $^6$Li+$X$:$^{87}$Rb$+X$.  This finding appears to be consistent with the universality hypothesis that asserts that the total collision rate coefficient for heavy collision partners is principally determined by the long-range van der Waals interactions \cite{booth2019}.  By contrast the K{\l}os and Tiesinga (KT) quantum scattering computation ratio predicts $\left< R^{\rm{KT}}_{7, 87} \right> = 0.770(3)$ and the BExp measurements yield $\left< R^{\rm{BExp}}_{7, 87} \right> = 0.756(21)$ for these same test species. The possible experimental sources of systematic error in our measurements have been examined in this paper, and we do not believe that they are large enough to explain these discrepancies.  In particular, we find that the deviation of the measured $\Gamma_{\rm{loss}}$ from the true $\Gamma_{\rm{tot}}$ due to the finite shallow trap depth and cold atom ensemble energy distributions is  $<0.5$\%, ensemble collision-induced heating uncertainties are less than 0.004\%, unaccounted 2- and 3-body intra-trap collisions $<0.05$\%, uncertainty associated with variation of the ambient temperature of the test environment $<0.012$\%, and drift of the background gas density $<0.14$\%.  

Because our systematic errors are so small and because the majority of the exprerimental values of $\svtot$ for $^{7}$Li collisions with test gases reported by Barker \textit{et al.}\cite{barker2023} are systematically 2-4\% higher than the theoretical predictions, we hypothesize  that the observed systematic differences in our $R_{6,87}$ values compared to the $R^{\rm{KT}}_{7,87}$ may arise from errors in the potential energy surfaces used in the KT quantum scattering computations for Li sensor atoms. Indeed a 3.4\% discrepancy is not large for such complex, multi-electron systems and underscores the difficulty of quantifying the uncertainties in \textit{ab initio} computations.  The agreement observed between the observed $R_X$, the \textit{ab initio} KT predictions\cite{klos2023} for H$_2$ may indicate that the potential energy surface computed for this much simpler system is more reliable.  We believe this work provides an exciting opportunity to help refine the theoretical models and associated experimental measurements for atom-based vacuum metrology.

\begin{acknowledgments}

We acknowledge financial support from the Natural Sciences and Engineering Research Council of Canada (NSERC/CRSNG) and the Canadian Foundation for Innovation (CFI).
This work was done at the Center for Research on Ultra-Cold Systems (CRUCS) and was supported, in part, through computational resources and services provided by Advanced Research Computing at the University of British Columbia.
\end{acknowledgments}

\section{Author Declarations}

\subsection{Conflict of Interest}

K.W.M and J.L.B. have
US patents 8,803,072 and 11,221,268 issued.

\subsection{Author Contributions}

{\bf Erik Frieling}: Conceptualization (equal); Data curation (equal); Formal analysis (equal); Investigation (lead); Methodology (equal); Project administration (equal); Software (equal); Visualization (lead); Writing – original draft (equal); Writing – review \& editing (equal).
{\bf Riley A.~Stewart}: Conceptualization (equal); Data curation (equal); Formal analysis (equal); Investigation (lead); Methodology (equal); Project administration (equal); Software (equal); Visualization (equal); Writing – original draft (equal); Writing – review \& editing (equal).
{\bf James L.~Booth}: Conceptualization (equal); Formal analysis (equal); Investigation (equal); Methodology (equal); Project administration (equal); Visualization (equal); Writing – original draft (equal); Writing – review \& editing (equal).
{\bf Kirk W.~Madison}: Conceptualization (equal); Funding acquisition (lead); Investigation (equal); Methodology (equal); Project administration (lead); Writing – original draft (equal); Writing – review \& editing (equal).

\subsection{Data Availability}
The data that support the findings of this study are available from the corresponding author upon reasonable request.

\begin{appendix}

    \section{Error Estimates}

    \subsection{Effect of The Sensor Ensemble Energy Distribution}
    \label{app:energy}
    
    Here we discuss the systematic deviation of the $R$ values arising from the finite trap depth and the sensor ensemble energy distribution.
    
    To measure the total collision rate between the sensor atom and the surrounding room-temperature gas, we monitor the loss of sensor atoms from a weakly-confining quadrupole magnetic trap. Due to the finite depth of the confining potential, a fraction of collisions will not be energetic enough to induce loss. For a sensor atom at zero energy, the loss rate $\Gamma_\mathrm{loss}$ deviates from the total collision rate $\Gamma_\mathrm{tot}$ by a trap-depth dependent function $f$,
    \begin{equation}
    \begin{aligned}
        \Gamma_\mathrm{loss}(U) &= n \svloss(U) \\
        &= \Gamma_\mathrm{tot} \left[1-f(U)\right]. \\
    \end{aligned}
    \label{eq:trapdepthdep}
    \end{equation}
    Here, $f(U)$ represents the fraction of collisions with the thermal gas that do not generate loss for a given trap depth $U$, which vanishes as the trap depth $U$ approaches zero. The variation of $f(U)$ on the trap depth depends on the details of the underlying interaction potential and distribution of collisional energies. Both semi-classical and \textit{ab-initio} calculations \cite{booth2019, klos2023} as well as experimental measurements \cite{booth2019,shen2020} have been performed to determine $f(U)$ for gas species typically found in vacuum systems.
    
\begin{figure}
        \centering
        \includegraphics[width=\linewidth]{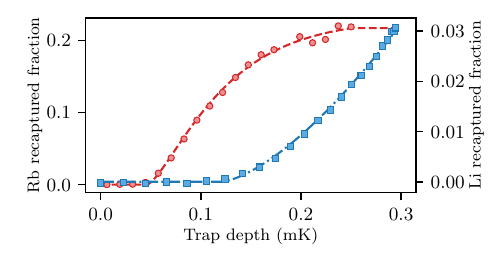}
        \caption{Experimentally measured recaptured fractions for Rb (red circles) and Li (blue squares) as a function of trap depth $U$ in the magnetic trap. The recaptured fraction error bars are smaller than the plotted points. For the Rb ensemble, the magnetic trap axial gradient is $B^\prime = 35$ G/cm, and the trap depth is determined by the minimum frequency in a time dependent radio-frequency field sweep as described previously.\cite{booth2019, stewart2022} For the Li ensemble, the axial gradient is $B^\prime = 100$ G/cm and the trap depth is limited by field dependence of the $\ket{1/2, -1/2}$ state. The dashed lines denote the fits to an offset Maxwell-Boltzmann energy distribution $\rho_\mathrm{MB}(E)$ described in Eq. \ref{eq:rhoMB}, and the corresponding parameters are listed in Table \ref{tab:energydist}. For the Li ensemble, we assume an estimated temperature of $T = 1.5(5)$ mK, as described previously.}
        \label{fig:ensembleenergies}
    \end{figure}

    In practice, the sensor ensemble contains atoms at non-zero energy, and the effective trap depth experienced by an atom with energy $E \geq 0 $ is given by $U-E$, corresponding to the energy transfer required to exceed the trap depth $U$. Consequently, the loss rate observed over the entire trapped ensemble is given by $\Gamma_\mathrm{loss}(U)$ averaged over the distribution of trapped energies $\rho(E)$, \cite{shen2021}
    \begin{equation}
    \begin{aligned}
        \overline{\Gamma_\mathrm{loss}(U)} &= n \overline{\langle \sigma_\mathrm{loss}(U) v \rangle}\\
        &= n \frac{\int_0^U \rho(E)\langle \sigma_\mathrm{loss}(U-E) v \rangle dE}{\int_0^U \rho(E) dE} \\
    \end{aligned}
    \label{eq:inteloss}
    \end{equation}
    Experimental methods for determining $\rho(E)$ have been described previously and entail measuring the number of sensor atoms in the trap below some energy $E$, providing the cumulative distribution function $F(U) = \int_{0}^{U} \rho(E) dE$.\cite{booth2019, stewart2022} Briefly, we measure the recaptured fraction $f_\mathrm{recap}$ after exposing the trapped ensemble to RF radiation that depletes the trapped population with energies $E$ above a chosen trap depth $U$, as described in the main text. By varying the chosen trap depth $U$, we map out the cumulative distribution function for the trapped ensemble,
    \begin{equation}
        f_\mathrm{recap}(U) = f_\mathrm{recap}(U_\mathrm{max}) F(U)
    \end{equation}
    where $U_\mathrm{max}$ denotes the recaptured fraction at the maximum trap depth, i.e. the recaptured fraction measured without exposure to RF radiation. Empirically, we find the trapped energy distributions $\rho(E)$ are reasonably well described by an offset Maxwell-Boltzmann energy distribution,
    \begin{equation}
        \rho_\mathrm{MB}(\epsilon, T_\mathrm{s}) = \Theta(\epsilon)\cdot 2 \sqrt{\frac{\epsilon}{\pi}} \left(\frac{1}{k_B T_\mathrm{S}}\right)^{3/2} \exp\left(-\frac{\epsilon}{k_B T_\mathrm{S}}\right).
    \label{eq:rhoMB}
    \end{equation}
    where $\epsilon = E-E_\mathrm{min}$, $T_\mathrm{S}$ is the sensor ensemble temperature, and $\Theta(\epsilon)$ is the Heaviside step function. This describes a distribution shifted by an amount $E_{\min}\geq 0$ below which no atoms are observed. The experimentally measured energy distributions for both ensembles are shown in Fig. \ref{fig:ensembleenergies}. In $^{6}$Li, the unresolved hyperfine structure of the excited state involved in the D2 ($^2S_{1/2} \rightarrow ^2P_{3/2}$) transition inhibits sub-Doppler cooling mechanisms based upon differential AC stark shifts and optical pumping between Zeeman sub-levels. As such, we find energy distribution of the lithium ensemble resembles $\rho_\mathrm{MB}$ with ($3k_B T/2 \gg U$) an average energy much larger the experimental trap depths used in this work. In this limit, we are unable to determine a reasonable estimate for the ensemble temperature $T$ using fits to eq. \ref{eq:rhoMB}. Without sub-Doppler cooling mechanisms, and for sufficiently weak magnetic trapping gradients, one expects the energy of the trapped ensemble to be dominated by the initial kinetic energy upon loading into the magnetic trap; correspondingly, the ensemble temperature in the magnetic trap approaches the temperature of the ensemble in the MOT. We estimate the lithium ensemble temperature to be $1.5(5)$ mK based upon previous studies in similar experimental systems. \cite{ridinger2011large,burchianti2014efficient} 
    
    \begin{table}[b]
    \caption{\label{tab:energydist}
    Experimental parameters describing the sensor atom energy distribution $\rho_\mathrm{MB}(E)$ as determined by fits to eq. \ref{eq:rhoMB}. The number in brackets denotes the statistical uncertainty. For the Li ensemble temperature, we assume an estimated temperature based upon previously measured $^6$Li MOT temperatures conducted on similar experimental systems.\cite{ridinger2011large,burchianti2014efficient}}
    \begin{tabular}{ccc}
    \toprule
    Species  & $T/k_B$ ($\mu\text{K}$) & $E_\mathrm{min}/k_B$ ($\mu\text{K}$) \\ \midrule
    Rb                      & 54.0(1)                  & 46.1(1)               \\
    Li                     &  $1500(500)$                  & 116.7(8)     \\         
    \bottomrule
    \end{tabular}
    \end{table}
    For the purposes of determining the ratio of total rate coefficients $R$, we measure the sensor atom loss rates at exceedingly shallow trap depths, where the loss rate $\overline{\Gamma_\mathrm{loss}(U)}$ approaches the total collision rate $\Gamma_\mathrm{tot}$. To characterize the deviation of the loss rate from the total collision rate due to the non-zero trap depth $U$ and finite-energy distribution $\rho(E)$ of the trapped ensemble, we use the trap-depth dependence model based upon \textit{ab-initio} calculations performed by K{\l}os and Tiesinga (KT). \cite{klos2023} We note that a trap-depth dependence model based on Quantum Diffractive Universality provides similar estimates for the expected deviation in the Rb loss rate. \cite{booth2019, shen2021}

    
    

    

    We calculate the deviation of the measured loss rate from the total collision rate by integrating $\langle \sigma_\mathrm{loss}(U) v \rangle$ over the experimentally-measured energy distributions, as in eq. \ref{eq:inteloss}, for both trapped species. From this, we compute the deviation in $R$ and present the results in table \ref{tab:finiteEnergy}. Across all gas species investigated here, the expected deviation $\delta R$ relative to the value of $R$ at zero trap depth, due the effects of a finite trap depth $U$ and energy distribution $\rho(E)$, is less than $0.51\%$. The largest deviations are associated with gas species where the loss rate decreases fastest with increasing trap depth. Generally, these correspond to species with a large polarizability and/or small collisional energies at room temperature.\cite{shen2023} 
    \begin{table}[b]
    \caption{\label{tab:finiteEnergy}
    Expected systematic deviations of the $R$ value due to finite trap depth and trapped ensemble energy distribution effects for various test species. We first find the deviation of the loss rate coefficient $\langle \sigma_\mathrm{loss} v \rangle (T, U)$ from the total collision rate coefficient $\svtot(T)$, given the trap depth and sensor atom energy distributions for both Li and Rb. These are used to construct the predicted deviation in the value of $R$, relative to the value at zero trap depth, denoted $\delta R/R$.}
    \begin{tabular}{ccc}
    \toprule
    Bk. species & \begin{tabular}[c]{@{}c@{}}$\delta \langle\sigma_\mathrm{loss} v \rangle / \svtot$\\(Rb/Li) [\%]\end{tabular} & $\delta R / R$ [\%]\\\midrule
    H$_2$          & -0.151/-0.017                                                   & 0.129            \\
    He          & -0.058/-0.005                                                   & 0.051           \\
    Ne          & -0.216/-0.023                                                   & 0.185            \\
    N$_2$          & -0.316/-0.033                                                   & 0.270            \\
    Ar          & -0.323/-0.035                                                   & 0.280            \\
    Kr          & -0.449/-0.049                                                   & 0.383            \\
    Xe          & -0.588/-0.064                                                   & 0.500    \\
    \bottomrule
    \end{tabular}
    
    \end{table}
    
    

    
    \subsection{Collision-induced heating}
    Collision events with the background thermal gas, which include the test gas species $X$, result in momentum exchange between the collision partners, thereby redistributing the energies of the trapped particles. This leads to heating of the sensor atom ensemble, modifying the energy distribution $\rho(E)$.\cite{booth2019, shen2023, stewart2022, Deshmukh2023,HeatingPaper} As the average loss rate over the entire ensemble depends on this distribution (eq. \ref{eq:inteloss}), heating modifies the observed loss rate.\cite{shen2023,stewart2022,Deshmukh2023,HeatingPaper} We note that, given the experimentally realized trap depth and energy distributions listed in table \ref{tab:energydist}, the fraction of collisions that do not result in loss is less than 0.6\% across all species, significantly suppressing heating effects. Previous studies accounted for heating by experimentally measuring the time-evolution of the trapped distribution. \cite{shen2023, stewart2022,Deshmukh2023,HeatingPaper}
    
    For this work, we estimate the residual effects of heating by utilizing the trap-depth dependence of the loss rate coefficient $\langle \sigma_\mathrm{loss} v\rangle (U)$ to determine the post-collisional energy probability density function $P_t$. Our formulation to account for heating follows a rigorous treatment discussed previously \cite{Deshmukh2023,HeatingPaper} and recently applied to correct the measurements of Barker \emph{et al.} \cite{Eckel2023}. In general, the post-collisional energy of a trapped sensor particle depends on the relative velocity and scattering angle of the collision event. However, in the limit where the distribution of relative velocities is dominated by the contribution from the collisional partner in the background thermal gas, $P_t$ is approximately independent of the sensor particle energy. For a background thermal gas $X$ composed of particles of mass $m_\mathrm{X}$ at temperature $T_\mathrm{X}$, corrections to this approximation are on the order of $\sqrt{m_\mathrm{bg} T_\mathrm{S}/m_\mathrm{S} T_\mathrm{bg}}$, which, across all trapped and test species partners investigated in this work, contribute corrections less than 1\%.\cite{HeatingPaper} These are neglected in the following analysis. We also neglect possible intra-trap thermalization collisions between trapped sensor atoms in the Rb ensemble owing to low intra-trap densities. \cite{barker2022precise} Such intra-trap thermalization collisions are forbidden for the cold $^6$Li ensemble.
    
    Under this assumption, we take the simplification of a sensor particle with zero initial energy. Then, we may utilize the trap depth dependence of $\langle \sigma_\mathrm{loss} v\rangle(U)$ to determine $P_t$. Following eq. \ref{eq:trapdepthdep},
    \begin{equation}
        \langle \sigma_\mathrm{loss} v\rangle (U) = \svtot \left[1 - f(U)\right]
    \end{equation}
    where, as mentioned previously, $f(U)$ represents the fraction of collisions between a trapped sensor particle with zero energy and a particle in the thermal background that does not generate loss at a given trap depth $U$. Phrased in terms of post-collisional energy probability density $P_t$, 
    \begin{equation}
        f(U) = \int_0^U P_t(E) dE.
    \end{equation}
    Hence, the derivative of $\langle \sigma_\mathrm{loss} v\rangle (U)$ yields $P_t$,
    \begin{equation}
        P_t(U) = -\frac{d}{dU} \frac{\langle \sigma_\mathrm{loss} v\rangle (U)}{\svtot}.
    \end{equation}
    In the above approximation, by neglecting the sensor atoms' contribution to the collisional energy, we have implicitly assumed that the transferred energy $E \geq 0$, and thereby have neglected `cooling' collisions for which the transferred energy is less than zero.\cite{HeatingPaper} 
    
    For sufficiently low trap depths, the energy distribution of the trapped ensemble can be approximated by the contributions from two sub-ensembles,
    \begin{equation}
        \rho(E,t) \simeq \rho_0(E,t) + \rho_1(E,t)
    \end{equation}
    where $\rho_0$ and $\rho_1$ correspond to trapped populations that have experienced 0 or 1 collision events respectively.  In principle, higher order terms exist but are negligible at the trap depths considered here. Neglecting these terms corresponds to the regime of sufficiently small trap depths which satisfy,
    \begin{equation}
        U \ll U_\mathrm{d},\qquad \qquad U_\mathrm{d} = \frac{4 \pi \hbar^2 v_p}{m_\mathrm{S} \svtot}
    \end{equation}
    where $U_\mathrm{d}$ is a characteristic energy scale describing the typical energy imparted to an initially at-rest sensor atoms following a collision event with a partner from the thermal gas. Here, $v_p = \sqrt{2 k_B T_\mathrm{X}/m_\mathrm{X}}$ is the most probable speed of the Maxwell-Boltzmann distributed background thermal gas composed of particles of mass $m_\mathrm{X}$ at a temperature $T_\mathrm{X}$.
    
    Both populations are depleted at a rate of $\Gamma$ due to background collisions, with the small fraction of heating collision events transferring particles from $\rho_0$ to $\rho_1$. The energy distribution of the heated sub-ensemble $\rho_1$ is the sub-ensemble $\rho_0$ modified by the post-collisional energy distribution $P_t$. If we consider the population in the heated sub-ensemble $\rho_1(E) dE$ at a given energy $E$, the populations in the unheated sub-ensemble $\rho_0$ that may contribute to $\rho_1(E) dE$ are those with energies $E^\prime \leq E$. For each population $\rho_0(E^\prime) dE^\prime$ in the unheated sub-ensemble, the probability that a given collision with the thermal background yields a post-collisional energy of $E$ is given by $P_t(E-E^\prime)$, that is, the probability that the collision transferred the energy difference between the initial and final states. Thus, each population $\rho_0(E^\prime) dE^\prime$ in the unheated-ensemble contributes to the population in the heated sub-ensemble $\rho_1(E) dE$ by an amount given by,
    \begin{equation}
        P_t(E^\prime -E) \rho_0(E^\prime, t) dE^\prime
    \end{equation}
    The sum of all contributions to the population $\rho_1(E^\prime) dE^\prime$ is given by integrating over all energies $0\leq  E^\prime \leq E$.
    \begin{equation}
        \rho_1(E,t) dE = \int_0^{E} P_t(E - E^\prime) \rho_0(E^\prime,t) dE^\prime.
    \end{equation}
    Following Deshmukh \emph{et al.}\cite{HeatingPaper}, the time evolution of the heated ensemble may be modelled as a Poisson process, with collision events occurring at a rate $\Gamma$. That is, the probability of the trapped sensor atom  has undergone $k$ collisions is given by,
    \begin{equation}
    \text{P($k$ collisions)} = \frac{e^{-\Gamma t}(\Gamma t)^k}{k!}.
    \end{equation}
    Considering only the two leading order sub-ensembles $\rho_0$ and $\rho_1$, the approximate time-evolution of the trapped distribution is therefore governed by,
    \begin{equation}
    \begin{aligned}
        \rho(E&,t) dE \simeq \\ 
        &e^{-\Gamma t }\left[\rho_0(E) + \Gamma t \int_0^{E} P_t(E - E^\prime) \rho_0(E^\prime) dE^\prime\right] dE.
    \end{aligned}
    \end{equation}
    Using the KT trap-depth dependence model, we compute time-evolution of the ensemble energy distribution, from which the cumulative population below a given trap depth $U$ can be determined as a function of time.
    
    In accordance with our exceedingly low trap depths (and hence small number of heating collisions), we find the loss rates $\overline{\Gamma_\mathrm{loss}}$ derived from a two-point measurement scheme, including ensemble heating, are modified by less than 0.004\% across all test species for the Rb ensemble, and less than 0.0004\% for the Li ensemble given our experimental conditions. Table \ref{tab:heating} presents the combined relative deviation $\delta R/R$ due to ensemble heating. Across all gas species, the expected deviations due to ensemble heating are less than 0.004\%.

    \begin{table}
    \caption{\label{tab:heating}
    Estimated systematic deviation due to ensemble heating for various gas species, given the experimental conditions given in table \ref{tab:energydist}.}
    \begin{tabular}{ccc}
    \toprule
    Bk. species & \begin{tabular}[c]{@{}c@{}}$\delta \overline{\Gamma_\mathrm{loss}} / \overline{\Gamma_\mathrm{loss}}$\\(Rb/Li) [$10^{-4}$\%]\end{tabular} & $\delta R / R$ [$10^{-4}$\%]\\\midrule
    H$_2$          & -2.43/-0.04                                                   & -2.39            \\
    He          & -0.357/-0.004                                                   & -0.353           \\
    Ne          & -5.01/-0.06                                                   & -4.95            \\
    N$_2$          & -10.43/-0.14                                                   & -10.29            \\
    Ar          & -11.28/-0.16                                                   & -11.12            \\
    Kr          & -21.12/-0.29                                                   & -20.83         \\
    Xe          & -36.17/-0.50                                                   & -35.67    \\
    \bottomrule
    \end{tabular}
    
    \end{table}

    \subsection{Density-dependent losses}
    \label{app:nonlinear}
    
\begin{figure}
        \centering
        \includegraphics[width=\linewidth]{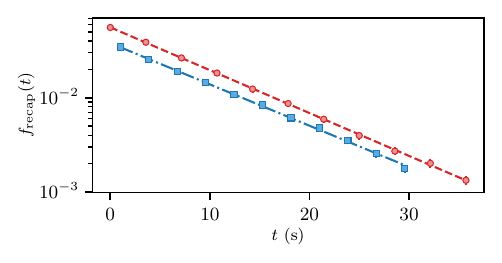}
        \caption{Typical atom loss curves for magnetically-trapped Rb (red circles) and Li (blue squares) ensembles subject to collision-induced loss due to a background thermal gas of Ne. The dashed and dash-dot curves denote fits to a one-body loss rate model providing $\Gamma^\mathrm{Rb}_\mathrm{loss} = 0.106(1)$ Hz and $\Gamma^\mathrm{Li}_\mathrm{loss} = 0.100(1)$ Hz for Rb or Li respectively. We observe no statistically-significant evidence of two-body or higher order loss.}
        \label{fig:example_decays}
    \end{figure}

    Trapped atoms are subject to a number of loss channels alongside to those generated by collisions with the thermal background gas. These include Majorana spin-flips, a process by which atoms undergo a non-adiabatic transition to an untrappable (high-field seeking) spin state.  In addition, there are higher $k$-body ($k \geq 2$) elastic and inelastic losses that depend on the intra-trap density. The differential atom loss per unit time is given by,
    \begin{equation}
        \frac{dn(\mathbf{r},t)}{dt} = -\Gamma n(\mathbf{r},t) - \beta n^2(\mathbf{r},t) + \mathcal{O}(n^3)
        \label{eq:atomloss}
    \end{equation}
    where $\Gamma$ and $\beta$ represents the one and two-body rate coefficients respectively, including inelastic and elastic losses. For typical intra-trap densities on the order of $10^{7}$ cm$^{-3}$, terms of order $n^3$ and above can be neglected. Example loss rate curves are shown in Fig.~\ref{fig:example_decays}; experimentally, we observe no statistically-significant evidence of two-body or higher order losses. 
    We also observe no statistically significant deviation between best-estimate values of $R$ using loss rate curves sampled with either linear spacing in time between $t =0 $ and $t = 1.5/\Gamma_\mathrm{i}$, and those taken using a two-point measurement scheme, for the same total number of recaptured fraction measurements. Both observations serve to support the validity of the two-point measurement scheme for determining the one-body loss rate. 
    
    We estimate an upper-bound for the systematic deviation in the inferred one-body loss rate measured from a two-point measurement scheme due to additional density-dependent losses by calculating the shift in the atom recapture due to this loss channel, constrained by our experimentally-measured loss rate curves. As our detection method measures total atom number $N = \int n ~d\mathbf{r}$, we integrate eq. \ref{eq:atomloss} over the spatial trapping volume to find,
    \begin{equation}
        \frac{dN}{dt} = -\Gamma N - \tilde{\beta} N^2
        \label{eq:dNdt}
    \end{equation}
    dropping higher-order terms, and where $\tilde{\beta}$ is given by,
    \begin{equation}
        \tilde{\beta}(t) = \beta \frac{\int_{\Omega} n^2(\mathbf{r},t) ~d\mathbf{r}}{\left( \int_\Omega n(\mathbf{r}, t) ~d\mathbf{r}\right)^2}
    \end{equation}
    where $\Omega$ denotes the energetically-accessible spatial volume for particles retained in the magnetic trap. We assume that $\tilde{\beta}(t) \simeq \tilde{\beta}(0)$ is approximately constant over the trapping duration $t$, under which the solution to eq \ref{eq:dNdt} is given by,
    \begin{equation}
        N(t) = \frac{N_0 e^{-\Gamma t}}{1 + \frac{\tilde{\beta} N_0}{\Gamma}(1-e^{-\Gamma t})}
        \label{eq:Ntconstprofile}
    \end{equation}
    where $N_0$ is the initial atom number at $t=0$. For the data presented in Fig. \ref{fig:example_decays}, we find $\tilde{\beta}_\mathrm{Rb} = -0.002(63)$ s$^{-1}$ and $\tilde{\beta}_\mathrm{Li} = -0.07(17)$ s$^{-1}$ for Rb and Li respectively, both statistically consistent with zero. We note that as $^6$Li is fermionic and the magnetically trapped atoms are in identical spin states, two-body collisions (and thus density-dependent losses) are forbidden for s-wave collisions and frozen out for p-wave collisions at the sensor-atom energies for these measurements.\cite{shen2023} We thus set $\tilde{\beta}_\mathrm{Li} = 0$ s$^{-1}$, and assume an upper bound of $\tilde{\beta}_\mathrm{Rb} \leq 0.061$ s$^{-1}$ for the Rb ensemble.

    In order to calculate an upper bound on the resulting shift in $R$, we perform a Monte-Carlo simulation of the sequence of measurements used to construct $R$. First, we draw 10 samples from a uniform distribution of background pressures corresponding to one-body loss rates between 0.07-0.4 s$^{-1}$, corresponding to a experimentally-relevant range of one-body loss rates (see Fig. \ref{fig:gvgplot}). At each pressure, the loss rate for the Rb and Li ensembles is calculated from the atom number at two trapped times $t = 0$ and $t = 1.5/\Gamma$, as determined by eq. \ref{eq:Ntconstprofile}, with the $\tilde{\beta}$ values fixed for each species to the values specified above. We then determine an inferred value of $R$, given the simulated loss rate measurements at the 10 randomly distributed test gas pressures, using ODR and a linear model for the relationship between $\Gamma_\mathrm{Li}$ and $\Gamma_\mathrm{Rb}$, as performed for the experimental measurements. This process is then repeated until we converge to an approximately constant distribution of $R$ values, corresponding to the possible $R$ values one would obtain for a given experimental measurement and the two-body loss rate coefficients $\tilde{\beta}$ for the two species. From this distribution, we observe a relative deviation in the average value of $R$ of less than -0.05\% due to two-body loss.
    
    \subsection{Test and Background gas temperature}
    
    The thermally-averaged total collision rate coefficient $\svtot$ depends on temperature of the thermal gas $X$ in the vicinity of the magnetically-trapped sensor atoms. Writing this quantity in terms of an integral over the speed distribution of impinging background particles,\cite{booth2019}
    \begin{equation}
        \svtot = 4\pi \int_0^\infty v^3 \rho_\mathrm{X}(v, T_\mathrm{X}, m_\mathrm{X}) \sigma_\mathrm{tot}(v) dv 
    \end{equation}
    where the speed of the thermal gas is assumed to be Maxwell-Boltzmann distributed,
    \begin{equation}
        \rho_\mathrm{X}(v, T_\mathrm{X}, m_\mathrm{X}) = \sqrt{\frac{2}{\pi}} \left(\frac{m_\mathrm{X}}{k_B T_\mathrm{X}}\right)^{3/2}\exp \left(-\frac{mv^2}{2k_B T_\mathrm{X}}\right)
    \label{eq:MBspeeddist}
    \end{equation}
    and $m_\mathrm{X}$ is the mass of the background particle. To estimate the magnitude of the expected uncertainty due to fluctuations in the background temperature, we approximate the leading-order dependence of $\svtot$ using the Jeffreys-Born approximation for the elastic scattering phase shifts, for which one can derive for a van-der Waals potential of the form $V(r) = -C_6/r^6$,\cite{booth2019}
    \begin{equation}
    \begin{aligned}
    \svtot_{C_6} &= 8.494636 \left( \frac{C_6}{\hbar v_p}\right)^{\frac{2}{5}} v_p + 7.19729 \left(\frac{\hbar}{\mu}\right) \left(\frac{C_6}{\hbar v_p}\right)^{\frac{1}{5}} \\ 
    &= \svtot_{0} + \svtot_1
    \end{aligned}
    \end{equation}
    where $v_p = \sqrt{2k_B T_\mathrm{X} / m_\mathrm{X}}$ is the most probable speed given the thermal gas distribution in eq. \ref{eq:MBspeeddist}. We thus find,
    \bea 
    \delta \svtot &=& \frac{3}{10}\frac{\delta T_\mathrm{X}}{T_\mathrm{X}} \svtot_0 - \frac{1}{10} \frac{\delta T_\mathrm{X}}{T_\mathrm{X}} \svtot_1  \nonumber \\
    &=& \frac{3}{10}\frac{\delta T_\mathrm{X}}{T_\mathrm{X}} \svtot - \frac{2}{5} \frac{\delta T_\mathrm{X}}{T_\mathrm{X}} \svtot_1 
    \eea
    The relative uncertainty in $\svtot$ is then given by,
    \bea 
    \frac{\delta \svtot}{\svtot} &=& \frac{3}{10}\frac{\delta T_\mathrm{X}}{T_\mathrm{X}} - \frac{2}{5}\frac{\delta T_\mathrm{X}}{T_\mathrm{X}}\frac{\svtot_1}{\svtot}.
    \eea
    From the above, one can construct the resulting uncertainty in $R$,
    \bea 
    \frac{\delta R}{R} &=& \frac{\delta \svtot_{\rm{Li}}}{\svtot_{\rm{Li}}} - \frac{\delta \svtot_{\rm{Rb}}}{\svtot_{\rm{Rb}}} \nonumber \\ 
    &=& \frac{2}{5}\frac{\delta T_\mathrm{X}}{T_\mathrm{X}} \left[\frac{\svtot_{\rm{Rb,1}}}{\svtot_{\rm{Rb}}}-\frac{\svtot_{\rm{Li,1}}}{\svtot_{\rm{Li}}} \right]
    \eea
    Given calculated $C_6$ values found in the literature (\cite{jiang2015, klos2023}) and an estimated maximum uncertainty in the test and background gas temperatures of $\delta T_\mathrm{X} = 2$ K, we find estimated relative uncertainty $\delta R / R$ due to fluctuations in the temperature. The results are presented in table \ref{tab:bktemp}. Across all species, the relative deviation $\delta R / R$ is less than 0.012\%. We note that the order of magnitude of the above predicted deviation is consistent those calculated using the KT model across all species. \cite{klos2023}
    
    \begin{table}[]
    \caption{\label{tab:bktemp}Estimated systematic deviation due to variations in the temperature of the thermal gas in the vicinity of the trapped sensor atoms. The values of $C_6$ are given in atomic units, where $E_h$ is the Hartree energy and $a_0$ is the bohr radius. Values for H$_2$ and N$_2$ taken from K{\l}os \textit{et al.}, with values for all other species taken from Jiang \textit{et al}.\cite{klos2023, jiang2015} 
    }
    \begin{tabular}{lll}
    \toprule
    Bk. species & \begin{tabular}[c]{@{}c@{}} $C_6$ ($E_h a_0^6$)\\(Rb/Li)\end{tabular} & $\delta R / R_0$ {[}\%{]} \\\midrule
    H$_2$         & $148.8$ / $82.8$ & 0.004                    \\
    He         & $44.7$ / $22.5$& 0.006                    \\
    Ne         & $88.2$ / $43.8$& 0.008                    \\
    N$_2$         & $349$ / $184$& 0.008                    \\
    Ar         & $337$ / $174$& 0.008                    \\
    Kr         & $499$ / $260$& 0.01                    \\
    Xe         & $782$ / $411$& 0.012      \\\bottomrule       
    \end{tabular}
    \end{table}

    \subsection{Contributions of other background gases}
    \label{app:gases}
    Throughout this work, we have assumed that the composition and density of the residual background gases other than the test gas remains static throughout the measurement.
    In this section, we relax this assumption and estimate the contribution to the uncertainty in $R_X$ due to changes in the background gas density.
    These changes can either be random shot-to-shot fluctuations, which contribute only to the statistical uncertainty, or changes where the density of the background gas is correlated to the test gas density, which contribute to the systematic uncertainty in $R$.
    
    In this section we estimate this systematic contribution.
    Correlated changes in the density of redisual background gases and the test gas could occur due to contamination of the test gas before it is leaked in or due to the changing gas load affecting the behaviour of the ion pumps, NEGs or the turbo pump. 
    We begin by deriving an expression for the systematic error in $R$ in terms of the ratio of the $R$ values for the test gas to the residual background gases, and the estimated correlated change in density at the location of the atoms. Then we explain how we use the RGA readings to estimate the density of the background and test gases at the location of the MT/MOT.

    
    We start by writing $R$ in terms of $\dni$, the difference between the maximum and minimum density of background gas $i$ over the course of a measurement of $R_X$ for gas $X$.
    This correlation would appear as a systematic rise and fall of the RGA trace for the background gas $i$ as the test gas density $n_X$ is changed as shown in Figs.~\ref{fig:gvgplot}b and \ref{fig:gvgplot}c.
    $R$ can be expressed as

    \begin{align}
    \begin{split}
        R &=
        \frac{\Delta\gli}{\Delta\grb} \\ &=
        \frac{\sum_{i} \dni \svtlii}{
            \sum_{i} \dni \svtrbi
        },
    \end{split}
    \end{align}
    where $\Delta \gli/\Delta \grb$ is the slope of the loss rate data as shown in Fig.~\ref{fig:gvgplot}a, which is potentially subject to the systematic uncertainty discussed here. 
    $\svtlii$ and $\svtrbi$ are the total collision rate coefficients for Li+$i$ and Rb+$i$ collisions, respectively.  Implicit is the assumption that every collision results in loss.
    
    
    
    
    
    Pulling out a factor of $R_X$ for test gas $X$ and expanding assuming a weak correlation $\dni/\dnx \ll 1$ yields

    \begin{align}
    \begin{split}
       R&= 
        R_X \left(
            1+\sum_{i\neq X}\left\{\frac{\ggli}{\gglix} 
            \right. \right.\\
             & \qquad\qquad \left. \left. - \frac{\ggrb}{\ggrbx} + \mathcal{O}\left[\left(\frac{\Delta n_i}{\Delta n_X}\right)^2\right]
        \right\}
        \right).
        \end{split}
    \end{align}
    
    We rewrite this in terms of the change in the partial pressure of each background species at the location of the sensor atoms $\Delta{P_i} = \Delta{n_i} k_B T$ and $R_i = \svtlii/\svtrbi$:
    
    \begin{align}
    R&= R_X + \delta_R^{(P)}, \\
    \delta_R^{(P)} &= R_X\left(\displaystyle\sum_{i\neq X}\left\{\left(\frac{\Delta{P_i}}{\Delta{P_X}}\right)\frac{\svtlii}{\svtli}{\left[1-\frac{R_X}{R_i}\right]}\right\}\right).
    \end{align}
    
    This analysis is limited to gases for which we know $R_i$. 
    Fortunately, the only mass numbers (apart from those corresponding to the test gas $X$) that show up in significant quantities on the RGA are 2 (H$_2$), 40 (Ar) and 28 (N$_2$), along with those corresponding to the test gas species we introduced.
    
    This analysis requires knowledge of the partial pressures of the gases at the location of the sensor atoms.  However, the RGA is not well calibrated to give quantitative measurements of the pressure, and there is differential pumping between the RGA and the location of the magnetic trap.  Thus, we start by estimating a correction factor $k_{X}^{\mathrm{RGA}} = \Delta P_X/\Delta P_{X}^{\mathrm{RGA}}$ relating a change in the partial pressure $\Delta P_X^{\mathrm{RGA}}$ of gas $X$ as measured by the RGA to a change in the partial pressure $\Delta P_X$ of the same gas at the location of the MOT/MT, where the latter is determined using the observed Li loss rate : 
    \begin{align}
        \Delta P_X = \frac{\Delta \gli}{\svtli k_B T},
    \end{align}
    where $\Delta \gli$ is the difference in the maximum and minimum loss rate of Li measured during a measurement of $R$ for a particular test gas and we use the $\svtli$ value from the quantum scattering calculations by K{\l}os and Tiesinga to convert the loss rate into an estimate of the pressure at the locations of the atoms.\cite{klos2023}  This derivation assumes that the test gas is the dominant contribution to the loss rate, which is justified for these measurements where the partial pressure of the test gas is significantly larger than all other gases in the vacuum.
    
    We calculate this correction factor $k_{X}^{ \mathrm{RGA}}$ for all 7 gases we investigated.
    Finally, we calculate 
    \begin{equation}
        \Delta P_i=k_{i}^{\mathrm{RGA}} \Delta P_{i}^{\mathrm{RGA}}
    \end{equation}
    for the residual background gases and estimate the contribution $\delta_R^{(P)}$, as shown in table \ref{tab:perror}.
    
    Since Helium has by far the lowest $R$ value, it provides a relatively large contribution to $\delta_R^P$.
    We include a column showing the values of $\delta_R^P$ with the contribution of Helium removed. 
    From table \ref{tab:perror}, it is clear that the contribution of these pressure fluctuations is well below the statistical error, $\le 0.8\%$. 
    
    \begin{table}
    \caption{\label{tab:perror} Estimated deviation $\delta_R^{(P)}/R$ of the measured $R$ values due to the change in the partial pressure of gases other than the test gas with pressure change $\Delta P_X$. The relative systematic error $\delta_R^{(P)}/R$ with the contribution from He removed is also shown, since this is the largest negative contribution due to the low value $R_\mathrm{Li+He}$. The deviations are well below the statistical error of $\le 0.8\%$.
    }
    \addtolength{\tabcolsep}{+5pt}
    \begin{tabular}{llcc}
    \toprule
    {Test} &                  $\Delta P_X$ & $\;\;$ $\delta_R^{(P)}/R$ $\;\;$  & $\delta_R^{(P)}/R$ (w/o He) \\
    gas &           [Torr]                 &    [\%]                          &     [\%]                                  \\
    \midrule
    H$_2$   &  $3.4\times10^{\text{-}9}$ &                     $-0.067$ &             $8.8\times10^{\text{-}4}$ \\
    He      &  $7.2\times10^{\text{-}9}$ &                      $0.002$ &                                 \\
    Ne      &  $1.0\times10^{\text{-}8}$ &                     $-0.023$ &                               $0.335$ \\
    N$_2$   &  $4.0\times10^{\text{-}9}$ &                     $-0.018$ &                               $0.021$ \\
    Ar      &  $8.1\times10^{\text{-}9}$ &                     $-0.009$ &                               $0.120$ \\
    Kr      &  $9.4\times10^{\text{-}9}$ &                     $-0.012$ &                               $0.062$ \\
    Xe      &  $9.5\times10^{\text{-}9}$ &                     $-0.008$ &                               $0.135$ \\
    \bottomrule
    \end{tabular}
    \addtolength{\tabcolsep}{-5pt}
    
    \end{table}

    \section{A discussion of the origins of the observed 3.4\% systematic discrepancies between $R_{6,87}$ and $R^{\rm{KT}}_{7,87}$}
    In this work we have observed that the $R_{6,87} = \svtlix/\svtrbs$ value is remarkably constant for heavier collision partners, $X$ = N$_2$, Ar, Kr, and Xe. Indeed, the corresponding $R^{\rm{KT}}_{7,87} = \svtot_{\rm{^7Li+X}}/\svtrbs$ based on the total collision rate coefficients computed by K{\l}os and Tiesinga (KT) also demonstrate this same behaviour. However, the average of the  $R_{6, 87}$ values for these four heavy collision partners, $\left<R_{6,87} \right> = 0.7965(9)$, is systematically 3.4(7)\% larger that the corresponding average value, $\left<R^{\rm{KT}}_{7,87} \right> = 0.770(5)$.
     The $R_{6,87}$ reported here have been corrected for finite trap depth effects (an effect that is less than 0.5\% in all cases), and the $R^{\rm{KT}}_{7,87}$ have been matched to our experimental ambient temperature following the KT prescription.\cite{klos2023} The remaining systematic measurement errors (type B), as shown in Appendix A, have been estimated to be less than 0.15\% in this work and the estimated statistical errors on our individual $R_{6,87}$ are less than 0.8\%. Thus, we face the conclusion that the \textit{systematic} 3.4\% discrepancy is significant and must be due to some other, unaccounted for source.
     
     This systematic discrepancy could be the result of systematic errors in the $C_6$ coefficients reported in the literature. For example, each of the Li+$X$ ($X$ = N$_2$, Ar, Kr, and Xe) $C_6$ coefficients could be systematically underestimated by approximately $8-10$\% or the Rb-$X$ $C_6$ coefficients could be systematically overestimated by $8-10$\%. This explanation seems unlikely. An intriguing possibility is that the discrepancies arise from systematic errors in the potential energy surfaces (PES) used in the quantum scattering computations (QSC) which are the basis of the $R^{\rm{KT}}_{7,87}$ estimates. Accurate determination of the interaction potential (especially at short range) is difficult, especially when the number of electrons in the collision complex is large, as it involves solving a computationally complex many-body quantum problem.  In many cases, exact diagonalization is not possible and approximations are employed which may not accurately reflect the true potential.
    
    To illustrate this point, it is helpful to examine the case of H$_2$ collisions with the sensor atoms.  Recent work investigated the variation of the \emph{ab initio} Li+H$_2$ potentials that resulted from various levels of approximation, allowing the authors to converge on a PES that had a small underlying uncertainty \cite{makrides2019}.  For this case, the corresponding quantum scattering computation value, $R^{\rm{KT}}_{7,87} = 0.82(3)$, agrees with our value here $R_{6,87} = 0.824(5)$ and with our previous measurement, $R^{\rm{SExp}}_{6,87} = 0.83(5)$.
    
    However, the next simplest collision partner to model, He, yielded  $R_{6,87} = 0.729(6)$ compared to  $R^{\rm{KT}}_{7,87} = 0.69(2)$, a $2\sigma$ discrepancy. The $R_{6,87}$ experimental result does, however, agree with the value deduced from the measurements by Barker \textit{et al.}, $R^{\rm{BExp}}_{7,87} = 0.73(2)$. 
    
    To provide some guidance on this interpretation (that the potential energy surfaces used may be inaccurate), we investigate how the changes to the potential may manifest in the loss rate measurements, and we test which sensor species is most sensitive to these changes.  As stated in the main text, the collision-induced trap loss rate is characterized by,
    \bea
    \Gamma &=& n_X \svtot 
    \label{eq:gammaappendix}
    \eea
    for shallow traps. Here $n_X$ is the density of test gas impinging on the trapped atoms, $v$ is the relative speed of the collisions, and $\sigma_{\rm{tot}}(v)$ is the speed-dependent total elastic collision cross-section. The brackets, $\left< \  \right>$, indicate that one measures the average loss rate of many collision partners striking the trapped sensor ensemble from different directions and with a Maxwell-Boltzmann speed distribution (MB) set by the ambient temperature of the apparatus and the mass of the impinging particles. Quantum scattering computations (QSC) can be used to determine the collision cross-section, $\sigma_{\rm{tot}}(v)$, provided sufficiently accurate potential energy surfaces, 
    are known \textit{a priori}. 
    
    In their simplest form the potential energy functions describing the collision dynamics consist of a short range core region valid for inter-species separations, $r < r_c$, smoothly connected to a long-range van der Waals attraction, $-C_6/r^6 - C_8/r^8 - C_{10}/r^{10}$ for $r>r_c$.
    \textit{Ab initio} calculations of the van der Waals coefficients, $C_n$, have been performed for a wide range of collision partners. \cite{mitroy2007, derevianko2010, klos2023, jiang2015} Of these sources only Derevianko\cite{derevianko2010} and KT\cite{klos2023} report an explicit uncertainty estimates for the $C_6$ dispersion coefficients. The Derevianko $C_6$ coefficient uncertainties are less than 0.6\% for Li+(Ar, Kr, Xe) and for Rb+(Ar, Kr, Xe) and KT report uncertainties in $C_6$ for Li+N$_2$ and Rb+N$_2$ of less than 0.2\%. Dervianko \textit{et al.} do not report values for the $C_8$ and $C_{10}$ dispersion coefficients and KT take these from Jiang\cite{jiang2015}.  The core portion of the potential is a challenge to model and, like the dispersion coefficients, estimating its accuracy is difficult. Indeed, in the absence of direct experimental data to compare against, the accuracy of model core potentials is very difficult to establish. Instead of the potential accuracy, the variations between the results of different computational models are usually reported. Thus, it is possible that the observed systematic difference between the computed $R^{\rm{KT}}_{7, 87}$ and the measured $R_{6,87}$ values is due to a systematic error in the core portion of the PES used in the theoretical predictions, where the effects of this error are different for the Rb+$X$ collisions and the Li+$X$ collisions. We hypothesize that these differences may arise from the effects of glory scattering on the $\svtot$

\begin{center}
    \begin{figure}[t]
       \centering
        \begin{subfigure}
            \centering
            \includegraphics[width=\columnwidth]{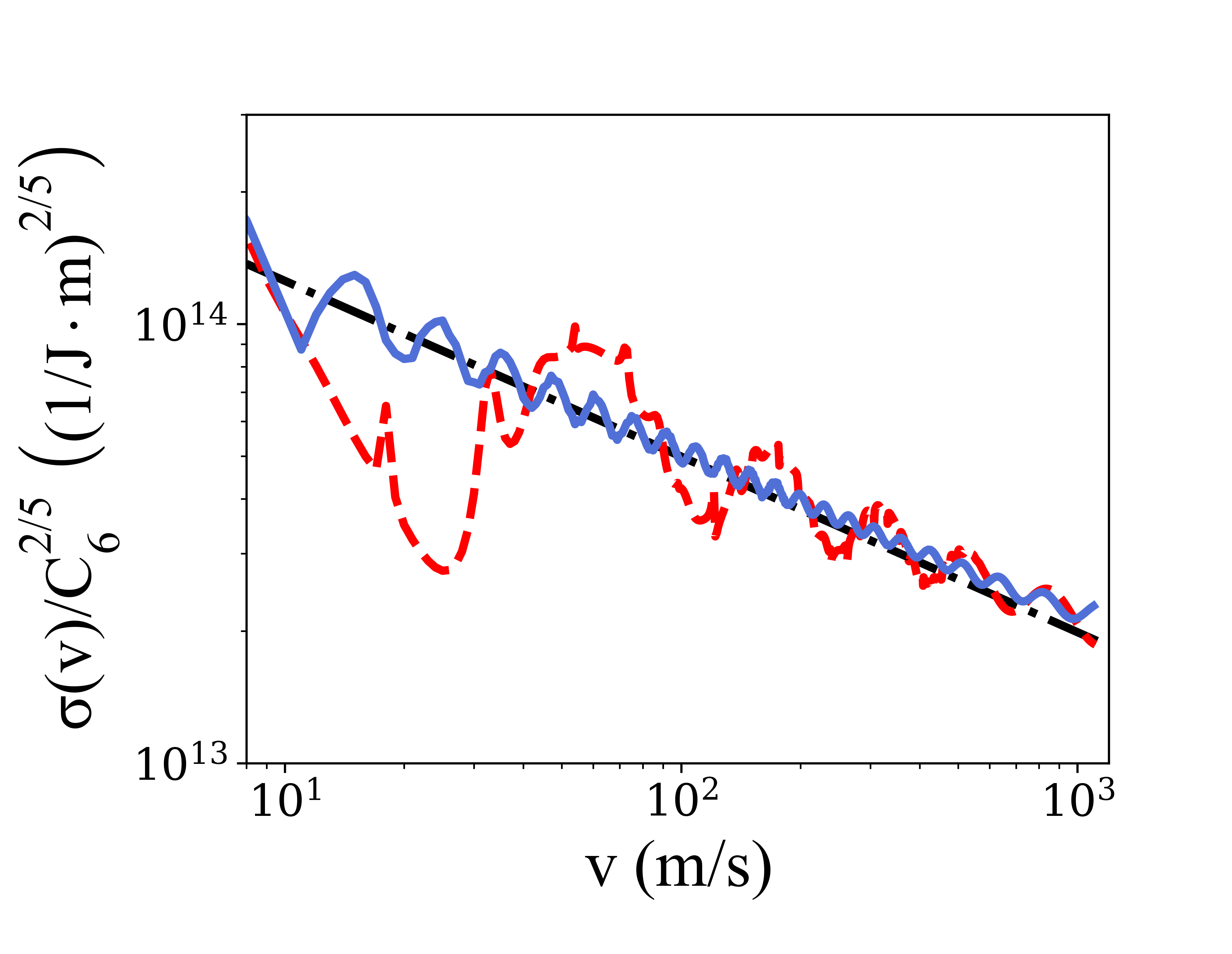} 
        \end{subfigure}
       \centering
        \begin{subfigure}
            \centering
            \includegraphics[width=\columnwidth]{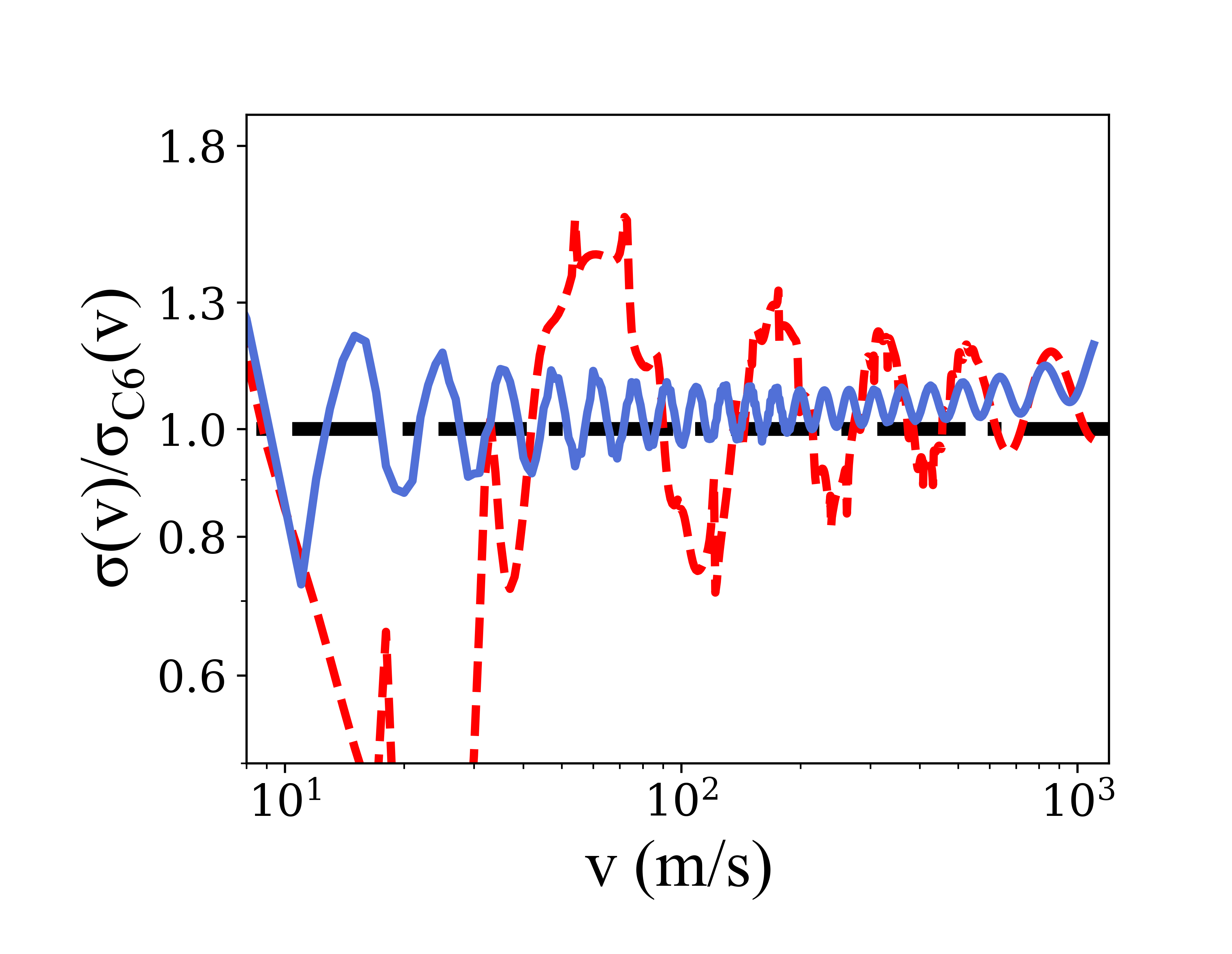} 
        \end{subfigure}
    \caption{Top panel: Plot of the normalized elastic cross-sections normalized, $\sigma(v)/C_6^{2/5}$ versus $v$ for Rb+Xe (blue solid trace), Li+Xe (red dashed trace) and the prediction from Eq.~\ref{eq:sigmavC6} for a purely $-C_6/r^6$ potential (black dash-dot trace). The computations, based on the interaction potentials published by Klos \textit{et al.} \cite{klos2023}, demonstrate the more pronounced and lower frequency glory oscillations for a light collision partner, Li, and compared to a heavy one, Rb, interacting with the same background gas species, Xe. \\
    Bottom panel: Plot of $\sigma(v)/\sigma_{\rm{C6}}(v)$ versus $v$. This normalization removes the purely $C_6$ long range portion of the computed cross-sections. The blue solid trace corresponds to Rb+Xe and the red dashed trace to Li+Xe. The presence of the long range $C_8$ and $C_{10}$ terms in the KT potentials used in the QSC lead to the upward trends at the high relative speed end of the plots.}
    \label{fig:gloryXe}
    \end{figure}
    \end{center}

    As stated above, the total collision rate coefficients for the heavier test gases Li+$X$ and Rb+$X$ ($X$ = N$_2$, Ar, Kr, and Xe), are remarkably constant. By their nature, the shallow trap loss rate measurements for these species are most sensitive to the long-range van der Waals-mediated, glancing collisions. Following Child\cite{child1974}, the elastic 
    collision cross-section between two species, $\sigma(v)$, is a smoothly decreasing cross-section with increasing relative velocity, $v$. For a simple long-range potential, $V(R) = -C_6/r^6$, the prediction is\cite{child1974},
    \bea
    \sigma_{\rm{C6}}(v)&\approx& 8.0828 \left(\frac{C_6}{\hbar v}\right)^{\frac{2}{5}}
    \label{eq:sigmavC6}
    \eea
    which leads to a cross-section decreasing as $v^{2/5}$.
    This picture is modified by the presence of so-called glory scattering (e.g. see Child\cite{child1974}): Some lower angular momentum (lower $L$) partial waves, associated with wavepackets which probe the core region of the PES,  scatter in the forward direction.  Which specific partial waves produce forward scattering is determined by the nature of the short- and long-range character of the potentials.  This physics is encoded in the elastic scattering phase shift for each wavepacket, $\eta((L, k)$, where $\hbar k$ is the magnitude of the momentum of the wavepacket.  Forward scattering occurs for the $L$ which satisfy, 
    \begin{equation}
    \frac{d\eta(L,k)}{dL} \approx 0
    \end{equation}
    for each $k$.
    These interfere with higher $L$ partial waves, associated with glancing collisions, which are also scattered in the forward direction. This results in the elastic collision cross-section, $\sigma(v)$, displaying interference oscillations about the smooth, decreasing cross-section, $\sigma_{\rm{C6}}(v)$. 
    Thus, the pattern of these glory oscillations depends on the details of the core portion of the potential describing the collision interaction. In general, the glory oscillations have the following characteristics \cite{child1974},
    \begin{enumerate}
    \item The number of oscillations per unit velocity decreases with a decreasing reduced mass of collision partners.
    \item The amplitude of the oscillations increases with a decreasing reduced mass of collision partners.
    \end{enumerate}
    These features are illustrated in Fig.~\ref{fig:gloryXe}, which shows scaled plots of quantum scattering computations (QSC) of $\sigma(v)$ for Li+Xe and Rb+Xe collisions, based on the interaction potentials published by K{\l}os \textit{et \ al} \cite{klos2023}.
    We observe the larger oscillation amplitude for the Li+Xe cross-sections compared to those for Rb+Xe, and the lower oscillation rate with speed for Li+Xe.
    
    One of the advantages of measuring  $\svtot$ is that the effects of these glory oscillations are at least partially mitigated by the averaging process. Namely,
    \bea
    \svtot &=& \int_0^\infty 4\pi v^2 dv \left(\rho_\mathrm{X}(v, T_\mathrm{X}, m_\mathrm{X}) \cdot \sigma(v)\cdot v\right) .
    \label{eq:svtotintegral}
    \eea
    Here $\rho_\mathrm{X}(v, T_\mathrm{X}, m_\mathrm{X})$ is the Maxwell-Boltzmann speed distribution of the test gas collision partners, Eq.~\ref{eq:MBspeeddist}. If there are numerous low-amplitude oscillations over the range of the integration, the glory oscillations, and, subsequently, the information they carry about the core potential, are suppressed. This leads to the universality of the $\svtot$ values discussed previously. \cite{booth2019, shen2020, shen2021}
    
\begin{figure}
       \centering
        \begin{subfigure}
            \centering
            \includegraphics[width=\columnwidth]{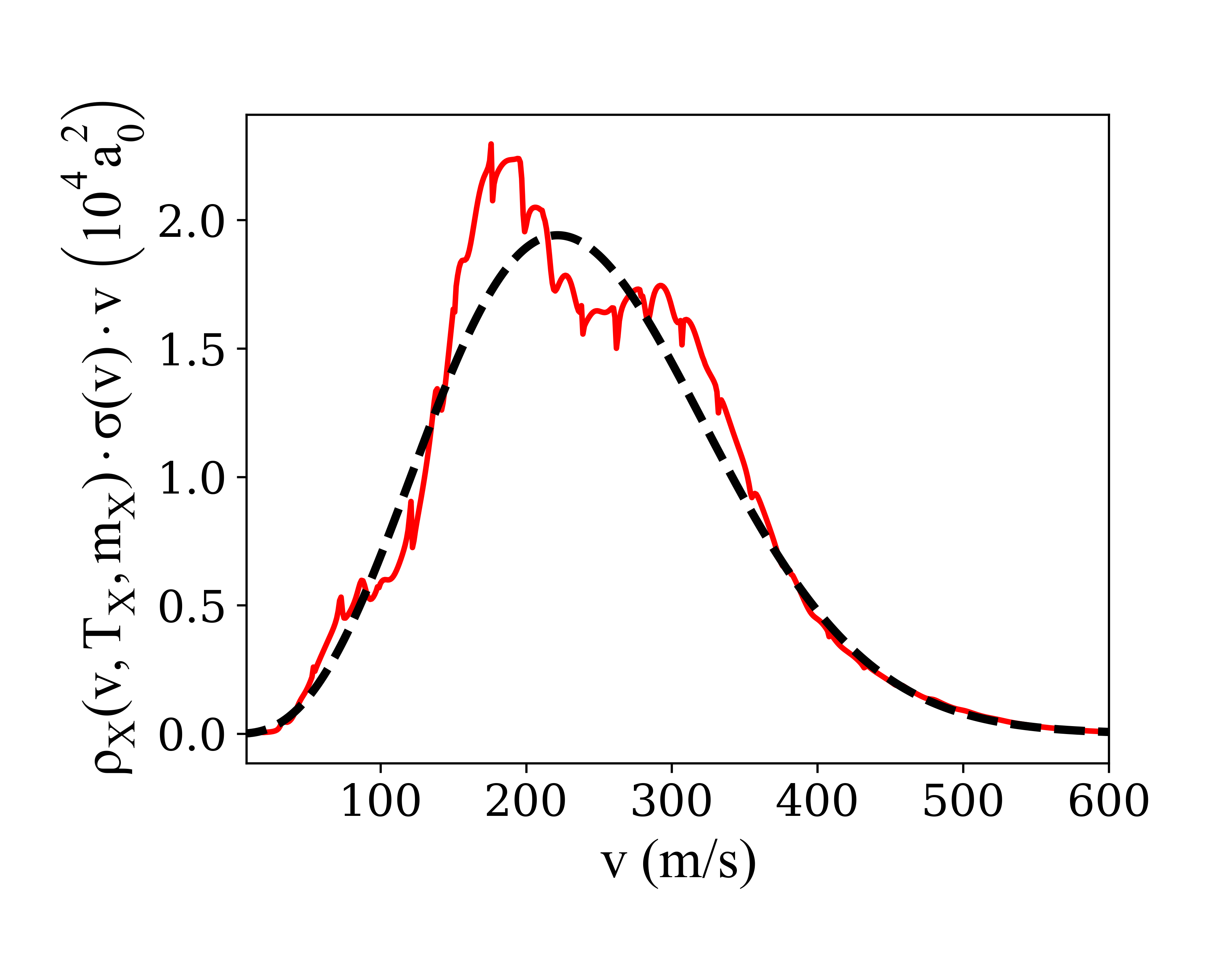} 
        \end{subfigure}
       \centering
        \begin{subfigure}
            \centering
            \includegraphics[width=\columnwidth]{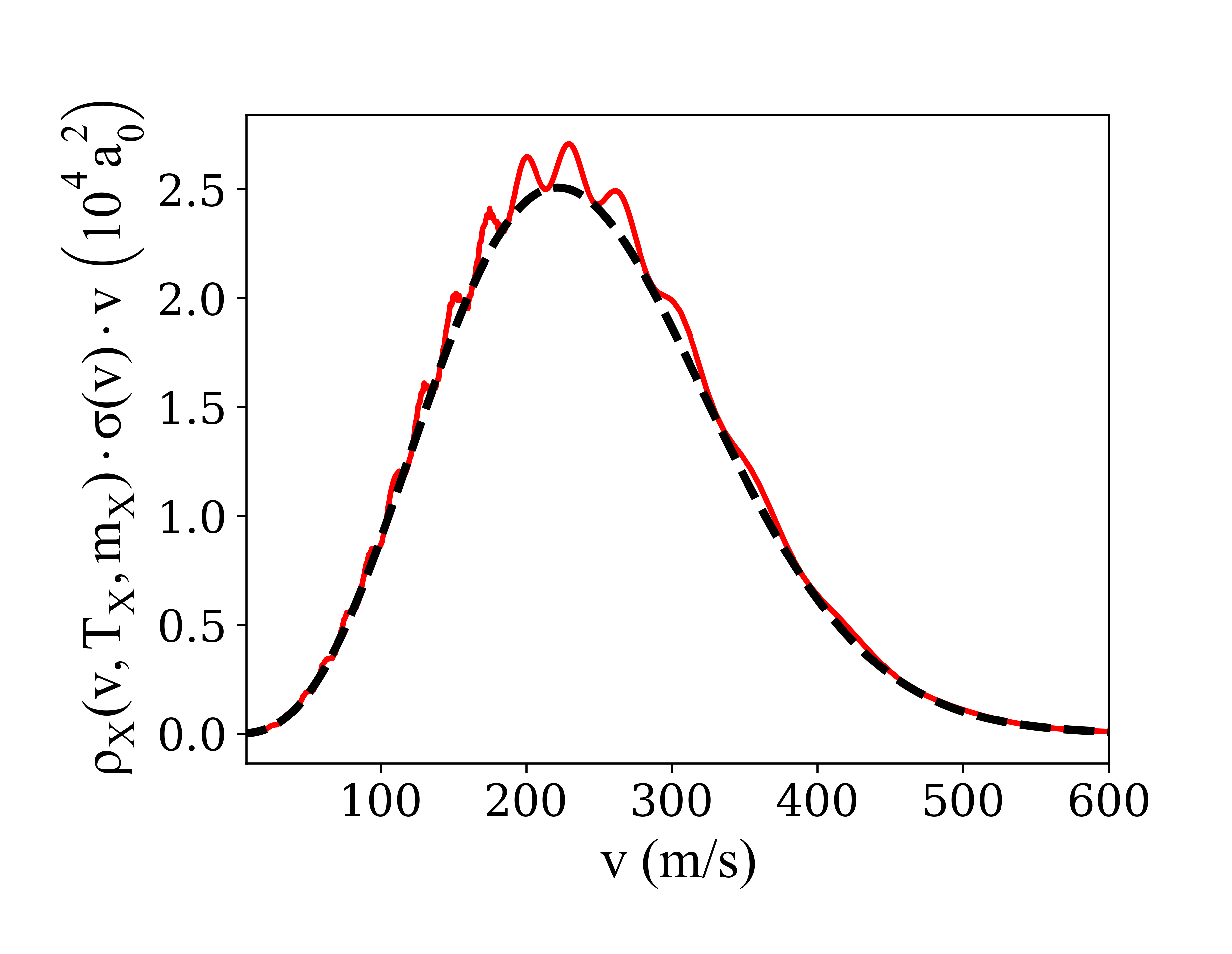} 
        \end{subfigure}
    \caption{The upper panel is a plot of the product $\rho_\mathrm{X}(v, T_\mathrm{X}, m_\mathrm{X})\cdot\sigma(v)\cdot v$ versus $v$ for Li+Xe while the lower panel is the corresponding plot for Rb+Xe. The solid red traces are the full quantum scattering computations (QSC) based on the KT potentials\cite{klos2023}, while the black dashed traces are the $\sigma_{C6}(v)$ predictions (Eq.~\ref{eq:sigmavC6}). The latter contain no glory oscillation effects and indicate a single term, purely the long-range interaction potential. For both of these plots the test gas temperature is 298 K.}
    \label{fig:MBvsigma}
    \end{figure}

    To underscore the difference in the behaviour between Li+Xe and Rb+Xe collisions, the products, $\left(\rho_\mathrm{X}(v, T_\mathrm{X}, m_\mathrm{X})~\cdot~\sigma(v)~\cdot~v\right)$, as a function of $v$ for these species are shown in Fig.~\ref{fig:MBvsigma}. One observes that the Rb+Xe plot approximates the ideal case with many small amplitude oscillation over the velocity range selected by the MB function. The Li+Xe plot, however, shows only a few large-amplitude oscillations over the same velocity range. Thus, one expects that the details of the core portion of the potential will not contribute significantly to $\svtot_{\rm{Rb+Xe}}$. By contrast, the details of the core will likely persist for the Li+Xe collisions. Thus, one might expect that changing the details of the core used in the QSC used to compute $\sigma(v)$ will lead to changes in the value of $\svtot$.
    \begin{table}
    \caption{\label{tab:QSC_SC}The QSC $\svtot_{\rm{KT}}$ values based on the potentials reported by \cite{klos2023} compared to SC computations, $\svtot_{\rm{SC}}$, at a temperature of 298 K. Column 4 shows the ratio of the SC to the QSC  $\svtot$. There is a systematic difference of 2\% between the QSC and SC values for the Li collisions compared to the near perfect agreement for the Rb computations. The statistical uncertainties (type A) for the $\svtot_{\rm{KT}}$ are less than 2\% and less than 0.3\% for the $\svtot_{\rm{SC}}$. (Note all $\svtot$ values have units of $\times 10^{-15}$ m$^3$/s.) }
    \centering
    \begin{tabular}{lccc}
    \toprule
     Species & $\svtot_{\rm{KT}}$ & $\svtot_{\rm{SC}}$  & $\svtot_{\rm{SC}}/\svtot_{\rm{KT}}$\\
    \hline
    Li+N$_2$&   2.65 & 2.70 &  1.019\\
    Li+Ar  &      2.34 & 2.37 & 1.015 \\
    Li+Kr &       2.15&  2.20 & 1.025 \\
    Li+Xe &       2.25&  2.30 & 1.025\\
    \hline
    Rb+N$_2$ & 3.46  & 3.49 & 1.007 \\
    Rb+Ar      & 3.045 & 3.035 & 0.997 \\
    Rb+Kr      & 2.80  & 2.79 & 0.999 \\
    Rb+Xe      &  2.88 & 2.89  & 1.003 \\
    \bottomrule  
    \end{tabular}
    
    \end{table}
    
    Thus, it is plausible that a large portion of the discrepancies between the measured values of $R_{6,87}$ presented here and the values derived from the theoretical QSC $\svtot$ of K{\l}os and Tiesinga\cite{klos2023} used in $R^{\rm{KT}}_{7,87}$ arise from details of the core of the interaction potentials between Li and its collision partners used in the theoretical estimates. While it is beyond the scope of this paper to investigate the detailed properties of the potentials, two tests were carried out.
    
    First, we validate the hypothesis that some of the information about the core repulsion persists for the Li+$X$ $\svtot$ while this information is absent in the corresponding Rb+$X$ $\svtot$. To illustrate this point, a semi-classical (SC) approximation was used which describes a solely long-range van der Waals interaction potential, $-C_6/r^6 - C_8/r^8 - C_{10}/r^{10}$. The corresponding Jeffreys-Born elastic scattering phase shifts for each $C_n$ term are\cite{child1974},
    \begin{equation}
    \eta_n(L, k) = -p(n) \left(\frac{\mu C_n}{\hbar^2}\right) \frac{k^{n-2}}{L^{n-1}}
    \label{eq:etaJB}
    \end{equation}
    where 
    \begin{equation}
    p(n) = \frac{1}{2}\Gamma\left(\frac{1}{2}\right)\cdot \frac{\Gamma\left(\frac{1}{2}\left(n-1\right)\right)}{\Gamma\left(\frac{1}{2}n\right)}.
    \end{equation}
    With this prescription the semi-classical (SC) total collision cross-sections, $\svtot_{\rm{SC}}$, were computed using the $C_n$ values published by KT\cite{klos2023}. These are compared to the corresponding KT QSC values, $\svtot_{\rm{KT}}$ in Table~\ref{tab:QSC_SC}. The fourth column shows the ratio of $\svtot_{\rm{SC}}:\svtot_{\rm{KT}}$. Since the SC description contains no core repulsion, the ratio is a measure of how thoroughly the velocity averaging suppresses the glory effects of the core on $\svtot$.   We see that the ratio is very close to 1 for the Rb+$X$ results, an indication that the glory oscillations have been strongly suppressed through the MB averaging. By contrast, the Li+$X$ SC values are \textit{systematically} $\approx 2$\% larger than the KT QSC values, indicating the persistence of some details of the core repulsion in the Li+$X$ computations (in this case reducing the KT $\svtot$ values through destructive interference). 
    
    The $\svtot$ values have statistical uncertainties (type A) associated with them: for the Li+$X$ $\svtot_{\rm{KT}}$ values these are less than 1\% for the collision partners listed in Table~\ref{tab:QSC_SC}. For Rb+$X$ ($X$ = Ar, Kr, or Xe) the statistical uncertainties on $\svtot_{\rm{KT}}$ are less than 0.4\% and 1.7\% for Rb+N$_2$. We approximate the corresponding uncertainty in $\svtot_{\rm{SC}}$ to be less than 0.25\% based on the statistical uncertainty equal to 0.6\% on the $C_6$ coefficients reported by Derevianko \cite{derevianko2010} for Li+$X$ and Rb+$X$. This leads to the estimate that the statistical uncertainties in the Li+$X$ $\svtot_{\rm{SC}}:\svtot_{\rm{KT}}$ ratio are $\le 1$\%. Thus, if the effects of glory scattering we removed by MB averaging for the ratio values should be distributed around 1.0, as seen for the Rb+$X$ results. By contrast, the Li+$X$  ratios $\svtot_{\rm{SC}}$:$\svtot_{\rm{KT}}$ show a \textit{systematic} 2\% offset, supporting the hypothesis that some information about the glory scattering is contained in the Li+$X$ $\svtot$ values.

    Next, the KT description of the interaction potential for Li+Xe\cite{klos2023} stitches together the short-range or core portion of the potential with the long-range van der Waals potential at an interspecies location labeled $r = r_c$. In KT's work\cite{klos2023}, $r_c = 18.5\ a_0$ for Li+Xe. Three different $r_c$ values were selected and the QSC for $\svtot$ were carried out to observe the shift in its value. The results are summarized in Table~\ref{tab:LiXeRc}.
    \begin{table}
    \centering
    \caption{\label{tab:LiXeRc}The variation in the Li+Xe $\svtot$ value at T = 298 K as a function of $r_c$, the location where the core and long-range van der Waals portions of the KT potentials connect. It is clear that the $\svtot$ values are not sensitive to this parameter. Neither the core potential nor the long range portion of the potential were varied in these computations.}
    
    \begin{tabular}{cc}
    \toprule
    $r_c$ $(a_0)$ & \phantom{aa} $\svtot$ ($\times 10^{-15}$m$^3$/s) \phantom{aa}\\
    \hline
    \phantom{(nomnom)}13.9 \phantom{(nom. value)}& 2.24 \\
    \phantom{(nomnom)}18.5 (nom. value) & 2.25 \\
    \phantom{(nomnom)}37.0 \phantom{(nom. value)}& 2.23 \\
    \bottomrule
    \end{tabular}
    
    \end{table}
    We observe variations of less than of 1\% even for very large changes in this parameter. indicating that shifting this parameter is not effective at changing the glory contributions to $\svtot$. 
    We believe these tests rule out the long-range portion of the potentials as being responsible for the discrepancies observed between our reported $R_{6,87}$ and $R^{\rm{KT}}_{7,87}$. 
    We hypothesize 
    that the core potentials described by K{\l}os \textit{et al.}\cite{klos2023} may need to be modified by adjusting the width and location of the potential minimum, and/or by adjusting the shape of the very short range repulsion curve to increase the computed $\svtot$ values. If this hypothesis is correct, then this provides a rare and exciting opportunity to investigate the shape of the short-range portion of the interaction potential experimentally.

\end{appendix}

\bibliography{CrossSpeciesCalibration}

\begin{thebibliography}{39}%
\makeatletter
\providecommand \@ifxundefined [1]{%
 \@ifx{#1\undefined}
}%
\providecommand \@ifnum [1]{%
 \ifnum #1\expandafter \@firstoftwo
 \else \expandafter \@secondoftwo
 \fi
}%
\providecommand \@ifx [1]{%
 \ifx #1\expandafter \@firstoftwo
 \else \expandafter \@secondoftwo
 \fi
}%
\providecommand \natexlab [1]{#1}%
\providecommand \enquote  [1]{``#1''}%
\providecommand \bibnamefont  [1]{#1}%
\providecommand \bibfnamefont [1]{#1}%
\providecommand \citenamefont [1]{#1}%
\providecommand \href@noop [0]{\@secondoftwo}%
\providecommand \href [0]{\begingroup \@sanitize@url \@href}%
\providecommand \@href[1]{\@@startlink{#1}\@@href}%
\providecommand \@@href[1]{\endgroup#1\@@endlink}%
\providecommand \@sanitize@url [0]{\catcode `\\12\catcode `\$12\catcode
  `\&12\catcode `\#12\catcode `\^12\catcode `\_12\catcode `\%12\relax}%
\providecommand \@@startlink[1]{}%
\providecommand \@@endlink[0]{}%
\providecommand \url  [0]{\begingroup\@sanitize@url \@url }%
\providecommand \@url [1]{\endgroup\@href {#1}{\urlprefix }}%
\providecommand \urlprefix  [0]{URL }%
\providecommand \Eprint [0]{\href }%
\providecommand \doibase [0]{https://doi.org/}%
\providecommand \selectlanguage [0]{\@gobble}%
\providecommand \bibinfo  [0]{\@secondoftwo}%
\providecommand \bibfield  [0]{\@secondoftwo}%
\providecommand \translation [1]{[#1]}%
\providecommand \BibitemOpen [0]{}%
\providecommand \bibitemStop [0]{}%
\providecommand \bibitemNoStop [0]{.\EOS\space}%
\providecommand \EOS [0]{\spacefactor3000\relax}%
\providecommand \BibitemShut  [1]{\csname bibitem#1\endcsname}%
\let\auto@bib@innerbib\@empty
\bibitem [{\citenamefont {Booth}\ \emph {et~al.}(2011)\citenamefont {Booth},
  \citenamefont {Fagnan}, \citenamefont {Klappauf}, \citenamefont {Madison},\
  and\ \citenamefont {Wang}}]{booth2011}%
  \BibitemOpen
  \bibfield  {author} {\bibinfo {author} {\bibfnamefont {J.}~\bibnamefont
  {Booth}}, \bibinfo {author} {\bibfnamefont {D.~E.}\ \bibnamefont {Fagnan}},
  \bibinfo {author} {\bibfnamefont {B.~G.}\ \bibnamefont {Klappauf}}, \bibinfo
  {author} {\bibfnamefont {K.~W.}\ \bibnamefont {Madison}},\ and\ \bibinfo
  {author} {\bibfnamefont {J.}~\bibnamefont {Wang}},\ }\href@noop {} {\bibinfo
  {title} {Method and device for accurately measuring the incident flux of
  amibient particles in a high or ultra-high vacuum environment}} (\bibinfo
  {year} {2011})\BibitemShut {NoStop}%
\bibitem [{\citenamefont {Madison}(2012)}]{madison2012}%
  \BibitemOpen
  \bibfield  {author} {\bibinfo {author} {\bibfnamefont {K.~W.}\ \bibnamefont
  {Madison}},\ }\href@noop {} {\bibinfo {title} {A cold atom based {{UHV}}
  pressure standard}},\ \bibinfo {howpublished} {personal communication}
  (\bibinfo {year} {2012})\BibitemShut {NoStop}%
\bibitem [{\citenamefont {Arpornthip}\ \emph {et~al.}(2012)\citenamefont
  {Arpornthip}, \citenamefont {Sackett},\ and\ \citenamefont
  {Hughes}}]{arpornthip2012}%
  \BibitemOpen
  \bibfield  {author} {\bibinfo {author} {\bibfnamefont {T.}~\bibnamefont
  {Arpornthip}}, \bibinfo {author} {\bibfnamefont {C.~A.}\ \bibnamefont
  {Sackett}},\ and\ \bibinfo {author} {\bibfnamefont {K.~J.}\ \bibnamefont
  {Hughes}},\ }\href@noop {} {\bibfield  {journal} {\bibinfo  {journal}
  {Physical Review A}\ }\textbf {\bibinfo {volume} {85}},\ \bibinfo {pages}
  {33420} (\bibinfo {year} {2012})}\BibitemShut {NoStop}%
\bibitem [{\citenamefont {Yuan}\ \emph {et~al.}(2013)\citenamefont {Yuan},
  \citenamefont {Ji}, \citenamefont {Zhao}, \citenamefont {Chang},
  \citenamefont {Xiao},\ and\ \citenamefont {Jia}}]{yuan2013}%
  \BibitemOpen
  \bibfield  {author} {\bibinfo {author} {\bibfnamefont {J.-P.}\ \bibnamefont
  {Yuan}}, \bibinfo {author} {\bibfnamefont {Z.-H.}\ \bibnamefont {Ji}},
  \bibinfo {author} {\bibfnamefont {Y.-T.}\ \bibnamefont {Zhao}}, \bibinfo
  {author} {\bibfnamefont {X.-F.}\ \bibnamefont {Chang}}, \bibinfo {author}
  {\bibfnamefont {L.-T.}\ \bibnamefont {Xiao}},\ and\ \bibinfo {author}
  {\bibfnamefont {S.-T.}\ \bibnamefont {Jia}},\ }\href
  {https://doi.org/10.1364/AO.52.006195} {\bibfield  {journal} {\bibinfo
  {journal} {Applied Optics}\ }\textbf {\bibinfo {volume} {52}},\ \bibinfo
  {pages} {6195} (\bibinfo {year} {2013})}\BibitemShut {NoStop}%
\bibitem [{\citenamefont {Moore}\ \emph {et~al.}(2015)\citenamefont {Moore},
  \citenamefont {Lee}, \citenamefont {Findlay}, \citenamefont
  {{Torralbo-Campo}}, \citenamefont {Bruce},\ and\ \citenamefont
  {Cassettari}}]{moore2015}%
  \BibitemOpen
  \bibfield  {author} {\bibinfo {author} {\bibfnamefont {R.~W.~G.}\
  \bibnamefont {Moore}}, \bibinfo {author} {\bibfnamefont {L.~A.}\ \bibnamefont
  {Lee}}, \bibinfo {author} {\bibfnamefont {E.~A.}\ \bibnamefont {Findlay}},
  \bibinfo {author} {\bibfnamefont {L.}~\bibnamefont {{Torralbo-Campo}}},
  \bibinfo {author} {\bibfnamefont {G.~D.}\ \bibnamefont {Bruce}},\ and\
  \bibinfo {author} {\bibfnamefont {D.}~\bibnamefont {Cassettari}},\ }\href
  {https://doi.org/10.1063/1.4928154} {\bibfield  {journal} {\bibinfo
  {journal} {Review of Scientific Instruments}\ }\textbf {\bibinfo {volume}
  {86}},\ \bibinfo {pages} {093108} (\bibinfo {year} {2015})}\BibitemShut
  {NoStop}%
\bibitem [{\citenamefont {Makhalov}\ \emph {et~al.}(2016)\citenamefont
  {Makhalov}, \citenamefont {Martiyanov},\ and\ \citenamefont
  {Turlapov}}]{makhalov2016}%
  \BibitemOpen
  \bibfield  {author} {\bibinfo {author} {\bibfnamefont {V.~B.}\ \bibnamefont
  {Makhalov}}, \bibinfo {author} {\bibfnamefont {K.~A.}\ \bibnamefont
  {Martiyanov}},\ and\ \bibinfo {author} {\bibfnamefont {A.~V.}\ \bibnamefont
  {Turlapov}},\ }\href@noop {} {\bibfield  {journal} {\bibinfo  {journal}
  {Metrologia}\ }\textbf {\bibinfo {volume} {53}},\ \bibinfo {pages} {1287}
  (\bibinfo {year} {2016})}\BibitemShut {NoStop}%
\bibitem [{\citenamefont {Makhalov}\ and\ \citenamefont
  {Turlapov}(2017)}]{makhalov2017}%
  \BibitemOpen
  \bibfield  {author} {\bibinfo {author} {\bibfnamefont {V.~B.}\ \bibnamefont
  {Makhalov}}\ and\ \bibinfo {author} {\bibfnamefont {A.~V.}\ \bibnamefont
  {Turlapov}},\ }\href {https://doi.org/10.1070/qel16353} {\bibfield  {journal}
  {\bibinfo  {journal} {Quantum Electronics}\ }\textbf {\bibinfo {volume}
  {47}},\ \bibinfo {pages} {431} (\bibinfo {year} {2017})}\BibitemShut
  {NoStop}%
\bibitem [{\citenamefont {Scherschligt}\ \emph {et~al.}(2017)\citenamefont
  {Scherschligt}, \citenamefont {Fedchak}, \citenamefont {Barker},
  \citenamefont {Eckel}, \citenamefont {Klimov}, \citenamefont {Makrides},\
  and\ \citenamefont {Tiesinga}}]{scherschligt2017}%
  \BibitemOpen
  \bibfield  {author} {\bibinfo {author} {\bibfnamefont {J.}~\bibnamefont
  {Scherschligt}}, \bibinfo {author} {\bibfnamefont {J.~A.}\ \bibnamefont
  {Fedchak}}, \bibinfo {author} {\bibfnamefont {D.~S.}\ \bibnamefont {Barker}},
  \bibinfo {author} {\bibfnamefont {S.}~\bibnamefont {Eckel}}, \bibinfo
  {author} {\bibfnamefont {N.}~\bibnamefont {Klimov}}, \bibinfo {author}
  {\bibfnamefont {C.}~\bibnamefont {Makrides}},\ and\ \bibinfo {author}
  {\bibfnamefont {E.}~\bibnamefont {Tiesinga}},\ }\href@noop {} {\bibfield
  {journal} {\bibinfo  {journal} {Metrologia}\ }\textbf {\bibinfo {volume}
  {54}},\ \bibinfo {pages} {S125} (\bibinfo {year} {2017})}\BibitemShut
  {NoStop}%
\bibitem [{\citenamefont {Scherschligt}\ \emph {et~al.}(2018)\citenamefont
  {Scherschligt}, \citenamefont {Fedchak}, \citenamefont {Ahmed}, \citenamefont
  {Barker}, \citenamefont {Douglass}, \citenamefont {Eckel}, \citenamefont
  {Hanson}, \citenamefont {Hendricks}, \citenamefont {Klimov}, \citenamefont
  {Purdy}, \citenamefont {Ricker}, \citenamefont {Singh},\ and\ \citenamefont
  {Stone}}]{scherschligt2018}%
  \BibitemOpen
  \bibfield  {author} {\bibinfo {author} {\bibfnamefont {J.}~\bibnamefont
  {Scherschligt}}, \bibinfo {author} {\bibfnamefont {J.~A.}\ \bibnamefont
  {Fedchak}}, \bibinfo {author} {\bibfnamefont {Z.}~\bibnamefont {Ahmed}},
  \bibinfo {author} {\bibfnamefont {D.~S.}\ \bibnamefont {Barker}}, \bibinfo
  {author} {\bibfnamefont {K.}~\bibnamefont {Douglass}}, \bibinfo {author}
  {\bibfnamefont {S.}~\bibnamefont {Eckel}}, \bibinfo {author} {\bibfnamefont
  {E.}~\bibnamefont {Hanson}}, \bibinfo {author} {\bibfnamefont
  {J.}~\bibnamefont {Hendricks}}, \bibinfo {author} {\bibfnamefont
  {N.}~\bibnamefont {Klimov}}, \bibinfo {author} {\bibfnamefont
  {T.}~\bibnamefont {Purdy}}, \bibinfo {author} {\bibfnamefont
  {J.}~\bibnamefont {Ricker}}, \bibinfo {author} {\bibfnamefont
  {R.}~\bibnamefont {Singh}},\ and\ \bibinfo {author} {\bibfnamefont
  {J.}~\bibnamefont {Stone}},\ }\href@noop {} {\bibfield  {journal} {\bibinfo
  {journal} {Journal of Vacuum Science \& Technology A}\ }\textbf {\bibinfo
  {volume} {36}},\ \bibinfo {pages} {040801} (\bibinfo {year}
  {2018})}\BibitemShut {NoStop}%
\bibitem [{\citenamefont {Xiang}\ \emph {et~al.}(2018)\citenamefont {Xiang},
  \citenamefont {Cheng}, \citenamefont {Peng}, \citenamefont {Wang},
  \citenamefont {Ren}, \citenamefont {Ji}, \citenamefont {Liu}, \citenamefont
  {Zhao}, \citenamefont {Li}, \citenamefont {Qu}, \citenamefont {Li},
  \citenamefont {Wang}, \citenamefont {Ye}, \citenamefont {Zhao}, \citenamefont
  {Yao}, \citenamefont {L{\"u}},\ and\ \citenamefont {Liu}}]{xiang2018}%
  \BibitemOpen
  \bibfield  {author} {\bibinfo {author} {\bibfnamefont {J.-f.}\ \bibnamefont
  {Xiang}}, \bibinfo {author} {\bibfnamefont {H.-n.}\ \bibnamefont {Cheng}},
  \bibinfo {author} {\bibfnamefont {X.-k.}\ \bibnamefont {Peng}}, \bibinfo
  {author} {\bibfnamefont {X.-w.}\ \bibnamefont {Wang}}, \bibinfo {author}
  {\bibfnamefont {W.}~\bibnamefont {Ren}}, \bibinfo {author} {\bibfnamefont
  {J.-w.}\ \bibnamefont {Ji}}, \bibinfo {author} {\bibfnamefont {K.-k.}\
  \bibnamefont {Liu}}, \bibinfo {author} {\bibfnamefont {J.-b.}\ \bibnamefont
  {Zhao}}, \bibinfo {author} {\bibfnamefont {L.}~\bibnamefont {Li}}, \bibinfo
  {author} {\bibfnamefont {Q.-z.}\ \bibnamefont {Qu}}, \bibinfo {author}
  {\bibfnamefont {T.}~\bibnamefont {Li}}, \bibinfo {author} {\bibfnamefont
  {B.}~\bibnamefont {Wang}}, \bibinfo {author} {\bibfnamefont {M.-f.}\
  \bibnamefont {Ye}}, \bibinfo {author} {\bibfnamefont {X.}~\bibnamefont
  {Zhao}}, \bibinfo {author} {\bibfnamefont {Y.-y.}\ \bibnamefont {Yao}},
  \bibinfo {author} {\bibfnamefont {D.-S.}\ \bibnamefont {L{\"u}}},\ and\
  \bibinfo {author} {\bibfnamefont {L.}~\bibnamefont {Liu}},\ }\href
  {https://doi.org/10.1088/1674-1056/27/7/073701} {\bibfield  {journal}
  {\bibinfo  {journal} {Chinese Physics B}\ }\textbf {\bibinfo {volume} {27}},\
  \bibinfo {pages} {073701} (\bibinfo {year} {2018})}\BibitemShut {NoStop}%
\bibitem [{\citenamefont {Booth}\ \emph {et~al.}(2019)\citenamefont {Booth},
  \citenamefont {Shen}, \citenamefont {Krems},\ and\ \citenamefont
  {Madison}}]{booth2019}%
  \BibitemOpen
  \bibfield  {author} {\bibinfo {author} {\bibfnamefont {J.~L.}\ \bibnamefont
  {Booth}}, \bibinfo {author} {\bibfnamefont {P.}~\bibnamefont {Shen}},
  \bibinfo {author} {\bibfnamefont {R.~V.}\ \bibnamefont {Krems}},\ and\
  \bibinfo {author} {\bibfnamefont {K.~W.}\ \bibnamefont {Madison}},\ }\href
  {https://doi.org/10.1088/1367-2630/ab452a} {\bibfield  {journal} {\bibinfo
  {journal} {New Journal of Physics}\ }\textbf {\bibinfo {volume} {21}},\
  \bibinfo {pages} {102001} (\bibinfo {year} {2019})}\BibitemShut {NoStop}%
\bibitem [{\citenamefont {Shen}\ \emph {et~al.}(2020)\citenamefont {Shen},
  \citenamefont {Madison},\ and\ \citenamefont {Booth}}]{shen2020}%
  \BibitemOpen
  \bibfield  {author} {\bibinfo {author} {\bibfnamefont {P.}~\bibnamefont
  {Shen}}, \bibinfo {author} {\bibfnamefont {K.~W.}\ \bibnamefont {Madison}},\
  and\ \bibinfo {author} {\bibfnamefont {J.~L.}\ \bibnamefont {Booth}},\ }\href
  {https://doi.org/10.1088/1681-7575/ab7170} {\bibfield  {journal} {\bibinfo
  {journal} {Metrologia}\ }\textbf {\bibinfo {volume} {57}},\ \bibinfo {pages}
  {025015} (\bibinfo {year} {2020})}\BibitemShut {NoStop}%
\bibitem [{\citenamefont {Shen}\ \emph {et~al.}(2021)\citenamefont {Shen},
  \citenamefont {Madison},\ and\ \citenamefont {Booth}}]{shen2021}%
  \BibitemOpen
  \bibfield  {author} {\bibinfo {author} {\bibfnamefont {P.}~\bibnamefont
  {Shen}}, \bibinfo {author} {\bibfnamefont {K.~W.}\ \bibnamefont {Madison}},\
  and\ \bibinfo {author} {\bibfnamefont {J.~L.}\ \bibnamefont {Booth}},\ }\href
  {https://doi.org/10.1088/1681-7575/abe02f} {\bibfield  {journal} {\bibinfo
  {journal} {Metrologia}\ }\textbf {\bibinfo {volume} {58}},\ \bibinfo {pages}
  {022101} (\bibinfo {year} {2021})}\BibitemShut {NoStop}%
\bibitem [{\citenamefont {Barker}\ \emph {et~al.}(2021)\citenamefont {Barker},
  \citenamefont {Klimov}, \citenamefont {Tiesinga}, \citenamefont {Fedchak},
  \citenamefont {Scherschligt},\ and\ \citenamefont {Eckel}}]{barker2021}%
  \BibitemOpen
  \bibfield  {author} {\bibinfo {author} {\bibfnamefont {D.}~\bibnamefont
  {Barker}}, \bibinfo {author} {\bibfnamefont {N.}~\bibnamefont {Klimov}},
  \bibinfo {author} {\bibfnamefont {E.}~\bibnamefont {Tiesinga}}, \bibinfo
  {author} {\bibfnamefont {J.}~\bibnamefont {Fedchak}}, \bibinfo {author}
  {\bibfnamefont {J.}~\bibnamefont {Scherschligt}},\ and\ \bibinfo {author}
  {\bibfnamefont {S.}~\bibnamefont {Eckel}},\ }\href
  {https://doi.org/10.1016/j.measen.2021.100229} {\bibfield  {journal}
  {\bibinfo  {journal} {Measurement: Sensors}\ }\textbf {\bibinfo {volume}
  {18}},\ \bibinfo {pages} {100229} (\bibinfo {year} {2021})}\BibitemShut
  {NoStop}%
\bibitem [{\citenamefont {Zhang}\ \emph {et~al.}(2022)\citenamefont {Zhang},
  \citenamefont {Sun}, \citenamefont {Dong}, \citenamefont {Wu}, \citenamefont
  {Li}, \citenamefont {Zhang}, \citenamefont {Zhang},\ and\ \citenamefont
  {Cheng}}]{zhang2022}%
  \BibitemOpen
  \bibfield  {author} {\bibinfo {author} {\bibfnamefont {S.-Z.}\ \bibnamefont
  {Zhang}}, \bibinfo {author} {\bibfnamefont {W.-J.}\ \bibnamefont {Sun}},
  \bibinfo {author} {\bibfnamefont {M.}~\bibnamefont {Dong}}, \bibinfo {author}
  {\bibfnamefont {H.-B.}\ \bibnamefont {Wu}}, \bibinfo {author} {\bibfnamefont
  {R.}~\bibnamefont {Li}}, \bibinfo {author} {\bibfnamefont {X.-J.}\
  \bibnamefont {Zhang}}, \bibinfo {author} {\bibfnamefont {J.-Y.}\ \bibnamefont
  {Zhang}},\ and\ \bibinfo {author} {\bibfnamefont {Y.-J.}\ \bibnamefont
  {Cheng}},\ }\href {https://doi.org/10.7498/aps.71.20212204} {\bibfield
  {journal} {\bibinfo  {journal} {Acta Phys. Sin.}\ }\textbf {\bibinfo {volume}
  {71}},\ \bibinfo {pages} {094204} (\bibinfo {year} {2022})}\BibitemShut
  {NoStop}%
\bibitem [{\citenamefont {Ehinger}\ \emph {et~al.}(2022)\citenamefont
  {Ehinger}, \citenamefont {Acharya}, \citenamefont {Barker}, \citenamefont
  {Fedchak}, \citenamefont {Scherschligt}, \citenamefont {Tiesinga},\ and\
  \citenamefont {Eckel}}]{ehinger2022}%
  \BibitemOpen
  \bibfield  {author} {\bibinfo {author} {\bibfnamefont {L.~H.}\ \bibnamefont
  {Ehinger}}, \bibinfo {author} {\bibfnamefont {B.~P.}\ \bibnamefont
  {Acharya}}, \bibinfo {author} {\bibfnamefont {D.~S.}\ \bibnamefont {Barker}},
  \bibinfo {author} {\bibfnamefont {J.~A.}\ \bibnamefont {Fedchak}}, \bibinfo
  {author} {\bibfnamefont {J.}~\bibnamefont {Scherschligt}}, \bibinfo {author}
  {\bibfnamefont {E.}~\bibnamefont {Tiesinga}},\ and\ \bibinfo {author}
  {\bibfnamefont {S.}~\bibnamefont {Eckel}},\ }\href
  {https://doi.org/10.1116/5.0095011} {\bibfield  {journal} {\bibinfo
  {journal} {AVS Quantum Science}\ }\textbf {\bibinfo {volume} {4}},\ \bibinfo
  {pages} {034403} (\bibinfo {year} {2022})}\BibitemShut {NoStop}%
\bibitem [{\citenamefont {Deshmukh}\ \emph {et~al.}(2023)\citenamefont
  {Deshmukh}, \citenamefont {Stewart}, \citenamefont {Shen}, \citenamefont
  {Booth},\ and\ \citenamefont {Madison}}]{HeatingPaper}%
  \BibitemOpen
  \bibfield  {author} {\bibinfo {author} {\bibfnamefont {A.}~\bibnamefont
  {Deshmukh}}, \bibinfo {author} {\bibfnamefont {R.~A.}\ \bibnamefont
  {Stewart}}, \bibinfo {author} {\bibfnamefont {P.}~\bibnamefont {Shen}},
  \bibinfo {author} {\bibfnamefont {J.~L.}\ \bibnamefont {Booth}},\ and\
  \bibinfo {author} {\bibfnamefont {K.~W.}\ \bibnamefont {Madison}},\
  }\href@noop {} {\bibinfo {title} {Trapped particle evolution driven by
  residual gas collisions}} (\bibinfo {year} {2023}),\ \Eprint
  {https://arxiv.org/abs/2310.04583} {arXiv:2310.04583 [physics.atom-ph]}
  \BibitemShut {NoStop}%
\bibitem [{\citenamefont {Makrides}\ \emph {et~al.}(2019)\citenamefont
  {Makrides}, \citenamefont {Barker}, \citenamefont {Fedchak}, \citenamefont
  {Scherschligt}, \citenamefont {Eckel},\ and\ \citenamefont
  {Tiesinga}}]{makrides2019}%
  \BibitemOpen
  \bibfield  {author} {\bibinfo {author} {\bibfnamefont {C.}~\bibnamefont
  {Makrides}}, \bibinfo {author} {\bibfnamefont {D.~S.}\ \bibnamefont
  {Barker}}, \bibinfo {author} {\bibfnamefont {J.~A.}\ \bibnamefont {Fedchak}},
  \bibinfo {author} {\bibfnamefont {J.}~\bibnamefont {Scherschligt}}, \bibinfo
  {author} {\bibfnamefont {S.}~\bibnamefont {Eckel}},\ and\ \bibinfo {author}
  {\bibfnamefont {E.}~\bibnamefont {Tiesinga}},\ }\href
  {https://doi.org/10.1103/PhysRevA.99.042704} {\bibfield  {journal} {\bibinfo
  {journal} {Physical Review A: Atomic, Molecular, and Optical Physics}\
  }\textbf {\bibinfo {volume} {99}},\ \bibinfo {pages} {042704} (\bibinfo
  {year} {2019})}\BibitemShut {NoStop}%
\bibitem [{\citenamefont {Makrides}\ \emph {et~al.}(2020)\citenamefont
  {Makrides}, \citenamefont {Barker}, \citenamefont {Fedchak}, \citenamefont
  {Scherschligt}, \citenamefont {Eckel},\ and\ \citenamefont
  {Tiesinga}}]{makrides2020}%
  \BibitemOpen
  \bibfield  {author} {\bibinfo {author} {\bibfnamefont {C.}~\bibnamefont
  {Makrides}}, \bibinfo {author} {\bibfnamefont {D.~S.}\ \bibnamefont
  {Barker}}, \bibinfo {author} {\bibfnamefont {J.~A.}\ \bibnamefont {Fedchak}},
  \bibinfo {author} {\bibfnamefont {J.}~\bibnamefont {Scherschligt}}, \bibinfo
  {author} {\bibfnamefont {S.}~\bibnamefont {Eckel}},\ and\ \bibinfo {author}
  {\bibfnamefont {E.}~\bibnamefont {Tiesinga}},\ }\href
  {https://doi.org/10.1103/PhysRevA.101.012702} {\bibfield  {journal} {\bibinfo
   {journal} {Physical Review A: Atomic, Molecular, and Optical Physics}\
  }\textbf {\bibinfo {volume} {101}},\ \bibinfo {pages} {012702} (\bibinfo
  {year} {2020})}\BibitemShut {NoStop}%
\bibitem [{\citenamefont {Makrides}\ \emph
  {et~al.}(2022{\natexlab{a}})\citenamefont {Makrides}, \citenamefont {Barker},
  \citenamefont {Fedchak}, \citenamefont {Scherschligt}, \citenamefont
  {Eckel},\ and\ \citenamefont {Tiesinga}}]{makrides2022}%
  \BibitemOpen
  \bibfield  {author} {\bibinfo {author} {\bibfnamefont {C.}~\bibnamefont
  {Makrides}}, \bibinfo {author} {\bibfnamefont {D.~S.}\ \bibnamefont
  {Barker}}, \bibinfo {author} {\bibfnamefont {J.~A.}\ \bibnamefont {Fedchak}},
  \bibinfo {author} {\bibfnamefont {J.}~\bibnamefont {Scherschligt}}, \bibinfo
  {author} {\bibfnamefont {S.}~\bibnamefont {Eckel}},\ and\ \bibinfo {author}
  {\bibfnamefont {E.}~\bibnamefont {Tiesinga}},\ }\href
  {https://doi.org/10.1103/PhysRevA.105.039903} {\bibfield  {journal} {\bibinfo
   {journal} {Physical Review A: Atomic, Molecular, and Optical Physics}\
  }\textbf {\bibinfo {volume} {105}},\ \bibinfo {pages} {039903} (\bibinfo
  {year} {2022}{\natexlab{a}})}\BibitemShut {NoStop}%
\bibitem [{\citenamefont {Makrides}\ \emph
  {et~al.}(2022{\natexlab{b}})\citenamefont {Makrides}, \citenamefont {Barker},
  \citenamefont {Fedchak}, \citenamefont {Scherschligt}, \citenamefont
  {Eckel},\ and\ \citenamefont {Tiesinga}}]{makrides2022a}%
  \BibitemOpen
  \bibfield  {author} {\bibinfo {author} {\bibfnamefont {C.}~\bibnamefont
  {Makrides}}, \bibinfo {author} {\bibfnamefont {D.~S.}\ \bibnamefont
  {Barker}}, \bibinfo {author} {\bibfnamefont {J.~A.}\ \bibnamefont {Fedchak}},
  \bibinfo {author} {\bibfnamefont {J.}~\bibnamefont {Scherschligt}}, \bibinfo
  {author} {\bibfnamefont {S.}~\bibnamefont {Eckel}},\ and\ \bibinfo {author}
  {\bibfnamefont {E.}~\bibnamefont {Tiesinga}},\ }\href
  {https://doi.org/10.1103/PhysRevA.105.029902} {\bibfield  {journal} {\bibinfo
   {journal} {Physical Review A: Atomic, Molecular, and Optical Physics}\
  }\textbf {\bibinfo {volume} {105}},\ \bibinfo {pages} {029902} (\bibinfo
  {year} {2022}{\natexlab{b}})}\BibitemShut {NoStop}%
\bibitem [{\citenamefont {K{\l}os}\ and\ \citenamefont
  {Tiesinga}(2023)}]{klos2023}%
  \BibitemOpen
  \bibfield  {author} {\bibinfo {author} {\bibfnamefont {J.}~\bibnamefont
  {K{\l}os}}\ and\ \bibinfo {author} {\bibfnamefont {E.}~\bibnamefont
  {Tiesinga}},\ }\href {https://doi.org/10.1063/5.0124062} {\bibfield
  {journal} {\bibinfo  {journal} {The Journal of Chemical Physics}\ }\textbf
  {\bibinfo {volume} {158}},\ \bibinfo {pages} {014308} (\bibinfo {year}
  {2023})}\BibitemShut {NoStop}%
\bibitem [{\citenamefont {Barker}\ \emph {et~al.}(2023)\citenamefont {Barker},
  \citenamefont {Fedchak}, \citenamefont {K{\l}os}, \citenamefont
  {Scherschligt}, \citenamefont {Sheikh}, \citenamefont {Tiesinga},\ and\
  \citenamefont {Eckel}}]{barker2023}%
  \BibitemOpen
  \bibfield  {author} {\bibinfo {author} {\bibfnamefont {D.~S.}\ \bibnamefont
  {Barker}}, \bibinfo {author} {\bibfnamefont {J.~A.}\ \bibnamefont {Fedchak}},
  \bibinfo {author} {\bibfnamefont {J.}~\bibnamefont {K{\l}os}}, \bibinfo
  {author} {\bibfnamefont {J.}~\bibnamefont {Scherschligt}}, \bibinfo {author}
  {\bibfnamefont {A.~A.}\ \bibnamefont {Sheikh}}, \bibinfo {author}
  {\bibfnamefont {E.}~\bibnamefont {Tiesinga}},\ and\ \bibinfo {author}
  {\bibfnamefont {S.~P.}\ \bibnamefont {Eckel}},\ }\href
  {https://doi.org/10.1116/5.0147686} {\bibfield  {journal} {\bibinfo
  {journal} {AVS Quantum Science}\ }\textbf {\bibinfo {volume} {5}},\ \bibinfo
  {pages} {035001} (\bibinfo {year} {2023})}\BibitemShut {NoStop}%
\bibitem [{\citenamefont {Eckel}(2023)}]{Eckel2023}%
  \BibitemOpen
  \bibfield  {author} {\bibinfo {author} {\bibfnamefont {S.~e.~a.}\
  \bibnamefont {Eckel}},\ }\href@noop {} {\bibinfo {title} {private
  communication}} (\bibinfo {year} {2023})\BibitemShut {NoStop}%
\bibitem [{\citenamefont {Madison}\ \emph {et~al.}(2018)\citenamefont
  {Madison}, \citenamefont {Booth}, \citenamefont {Shen},\ and\ \citenamefont
  {Krems}}]{madison2018}%
  \BibitemOpen
  \bibfield  {author} {\bibinfo {author} {\bibfnamefont {K.}~\bibnamefont
  {Madison}}, \bibinfo {author} {\bibfnamefont {J.}~\bibnamefont {Booth}},
  \bibinfo {author} {\bibfnamefont {P.}~\bibnamefont {Shen}},\ and\ \bibinfo
  {author} {\bibfnamefont {R.}~\bibnamefont {Krems}},\ }\href@noop {} {\bibinfo
  {title} {Quantum pressure standard and methods for determining and using
  same}} (\bibinfo {year} {2018})\BibitemShut {NoStop}%
\bibitem [{\citenamefont {Stewart}\ \emph {et~al.}(2022)\citenamefont
  {Stewart}, \citenamefont {Shen}, \citenamefont {Booth},\ and\ \citenamefont
  {Madison}}]{stewart2022}%
  \BibitemOpen
  \bibfield  {author} {\bibinfo {author} {\bibfnamefont {R.~A.}\ \bibnamefont
  {Stewart}}, \bibinfo {author} {\bibfnamefont {P.}~\bibnamefont {Shen}},
  \bibinfo {author} {\bibfnamefont {J.~L.}\ \bibnamefont {Booth}},\ and\
  \bibinfo {author} {\bibfnamefont {K.~W.}\ \bibnamefont {Madison}},\ }\href
  {https://doi.org/10.1103/PhysRevA.106.052812} {\bibfield  {journal} {\bibinfo
   {journal} {Physical Review A: Atomic, Molecular, and Optical Physics}\
  }\textbf {\bibinfo {volume} {106}},\ \bibinfo {pages} {052812} (\bibinfo
  {year} {2022})}\BibitemShut {NoStop}%
\bibitem [{\citenamefont {Shen}\ \emph {et~al.}(2023)\citenamefont {Shen},
  \citenamefont {Frieling}, \citenamefont {Herperger}, \citenamefont {Uhland},
  \citenamefont {Stewart}, \citenamefont {Deshmukh}, \citenamefont {Krems},
  \citenamefont {Booth},\ and\ \citenamefont {Madison}}]{shen2023}%
  \BibitemOpen
  \bibfield  {author} {\bibinfo {author} {\bibfnamefont {P.}~\bibnamefont
  {Shen}}, \bibinfo {author} {\bibfnamefont {E.}~\bibnamefont {Frieling}},
  \bibinfo {author} {\bibfnamefont {K.~R.}\ \bibnamefont {Herperger}}, \bibinfo
  {author} {\bibfnamefont {D.}~\bibnamefont {Uhland}}, \bibinfo {author}
  {\bibfnamefont {R.~A.}\ \bibnamefont {Stewart}}, \bibinfo {author}
  {\bibfnamefont {A.}~\bibnamefont {Deshmukh}}, \bibinfo {author}
  {\bibfnamefont {R.~V.}\ \bibnamefont {Krems}}, \bibinfo {author}
  {\bibfnamefont {J.~L.}\ \bibnamefont {Booth}},\ and\ \bibinfo {author}
  {\bibfnamefont {K.~W.}\ \bibnamefont {Madison}},\ }\href@noop {} {\bibfield
  {journal} {\bibinfo  {journal} {New Journal of Physics}\ } (\bibinfo {year}
  {2023})}\BibitemShut {NoStop}%
\bibitem [{\citenamefont {Bowden}\ \emph {et~al.}(2016)\citenamefont {Bowden},
  \citenamefont {Gunton}, \citenamefont {Semczuk}, \citenamefont {Dare},\ and\
  \citenamefont {Madison}}]{bowden2016}%
  \BibitemOpen
  \bibfield  {author} {\bibinfo {author} {\bibfnamefont {W.}~\bibnamefont
  {Bowden}}, \bibinfo {author} {\bibfnamefont {W.}~\bibnamefont {Gunton}},
  \bibinfo {author} {\bibfnamefont {M.}~\bibnamefont {Semczuk}}, \bibinfo
  {author} {\bibfnamefont {K.}~\bibnamefont {Dare}},\ and\ \bibinfo {author}
  {\bibfnamefont {K.~W.}\ \bibnamefont {Madison}},\ }\bibfield  {journal}
  {\bibinfo  {journal} {Review of Scientific Instruments}\ }\textbf {\bibinfo
  {volume} {87}},\ \href {https://doi.org/10.1063/1.4945567}
  {10.1063/1.4945567} (\bibinfo {year} {2016})\BibitemShut {NoStop}%
\bibitem [{\citenamefont {Bowden}(2014)}]{bowden2014}%
  \BibitemOpen
  \bibfield  {author} {\bibinfo {author} {\bibfnamefont {W.~J.}\ \bibnamefont
  {Bowden}},\ }\emph {\bibinfo {title} {An Experimental Apparatus for the Laser
  Cooling of Lithium and Rubidium}},\ \href
  {https://doi.org/10.14288/1.0167633} {Ph.D. thesis},\ \bibinfo  {school}
  {University of British Columbia} (\bibinfo {year} {2014})\BibitemShut
  {NoStop}%
\bibitem [{\citenamefont {Jousten}(2016)}]{jousten2016}%
  \BibitemOpen
  \bibfield  {author} {\bibinfo {author} {\bibfnamefont {K.}~\bibnamefont
  {Jousten}},\ }in\ \href {https://doi.org/10.1002/9783527688265.ch11} {\emph
  {\bibinfo {booktitle} {Handbook of {{Vacuum Technology}}}}}\ (\bibinfo
  {publisher} {{John Wiley \& Sons, Ltd}},\ \bibinfo {year} {2016})\
  Chap.~\bibinfo {chapter} {11}, pp.\ \bibinfo {pages} {463--510}\BibitemShut
  {NoStop}%
\bibitem [{\citenamefont {Jiang}\ \emph {et~al.}(2015)\citenamefont {Jiang},
  \citenamefont {Mitroy}, \citenamefont {Cheng},\ and\ \citenamefont
  {Bromley}}]{jiang2015}%
  \BibitemOpen
  \bibfield  {author} {\bibinfo {author} {\bibfnamefont {J.}~\bibnamefont
  {Jiang}}, \bibinfo {author} {\bibfnamefont {J.}~\bibnamefont {Mitroy}},
  \bibinfo {author} {\bibfnamefont {Y.}~\bibnamefont {Cheng}},\ and\ \bibinfo
  {author} {\bibfnamefont {M.~W.}\ \bibnamefont {Bromley}},\ }\href@noop {}
  {\bibfield  {journal} {\bibinfo  {journal} {Atomic Data and Nuclear Data
  Tables}\ }\textbf {\bibinfo {volume} {101}},\ \bibinfo {pages} {158}
  (\bibinfo {year} {2015})}\BibitemShut {NoStop}%
\bibitem [{\citenamefont {Tom}\ \emph {et~al.}(2009)\citenamefont {Tom},
  \citenamefont {Bhasker}, \citenamefont {Miyamoto}, \citenamefont {Momose},\
  and\ \citenamefont {McCall}}]{10.1063/1.3072881}%
  \BibitemOpen
  \bibfield  {author} {\bibinfo {author} {\bibfnamefont {B.~A.}\ \bibnamefont
  {Tom}}, \bibinfo {author} {\bibfnamefont {S.}~\bibnamefont {Bhasker}},
  \bibinfo {author} {\bibfnamefont {Y.}~\bibnamefont {Miyamoto}}, \bibinfo
  {author} {\bibfnamefont {T.}~\bibnamefont {Momose}},\ and\ \bibinfo {author}
  {\bibfnamefont {B.~J.}\ \bibnamefont {McCall}},\ }\href
  {https://doi.org/10.1063/1.3072881} {\bibfield  {journal} {\bibinfo
  {journal} {Review of Scientific Instruments}\ }\textbf {\bibinfo {volume}
  {80}},\ \bibinfo {pages} {016108} (\bibinfo {year} {2009})},\ \Eprint
  {https://arxiv.org/abs/https://pubs.aip.org/aip/rsi/article-pdf/doi/10.1063/1.3072881/14827893/016108\_1\_online.pdf}
  {https://pubs.aip.org/aip/rsi/article-pdf/doi/10.1063/1.3072881/14827893/016108\_1\_online.pdf}
  \BibitemShut {NoStop}%
\bibitem [{\citenamefont {Ridinger}\ \emph {et~al.}(2011)\citenamefont
  {Ridinger}, \citenamefont {Chaudhuri}, \citenamefont {Salez}, \citenamefont
  {Eismann}, \citenamefont {Fernandes}, \citenamefont {Magalhaes},
  \citenamefont {Wilkowski}, \citenamefont {Salomon},\ and\ \citenamefont
  {Chevy}}]{ridinger2011large}%
  \BibitemOpen
  \bibfield  {author} {\bibinfo {author} {\bibfnamefont {A.}~\bibnamefont
  {Ridinger}}, \bibinfo {author} {\bibfnamefont {S.}~\bibnamefont {Chaudhuri}},
  \bibinfo {author} {\bibfnamefont {T.}~\bibnamefont {Salez}}, \bibinfo
  {author} {\bibfnamefont {U.}~\bibnamefont {Eismann}}, \bibinfo {author}
  {\bibfnamefont {D.~R.}\ \bibnamefont {Fernandes}}, \bibinfo {author}
  {\bibfnamefont {K.}~\bibnamefont {Magalhaes}}, \bibinfo {author}
  {\bibfnamefont {D.}~\bibnamefont {Wilkowski}}, \bibinfo {author}
  {\bibfnamefont {C.}~\bibnamefont {Salomon}},\ and\ \bibinfo {author}
  {\bibfnamefont {F.}~\bibnamefont {Chevy}},\ }\href@noop {} {\bibfield
  {journal} {\bibinfo  {journal} {The European Physical Journal D}\ }\textbf
  {\bibinfo {volume} {65}},\ \bibinfo {pages} {223} (\bibinfo {year}
  {2011})}\BibitemShut {NoStop}%
\bibitem [{\citenamefont {Burchianti}\ \emph {et~al.}(2014)\citenamefont
  {Burchianti}, \citenamefont {Valtolina}, \citenamefont {Seman}, \citenamefont
  {Pace}, \citenamefont {De~Pas}, \citenamefont {Inguscio}, \citenamefont
  {Zaccanti},\ and\ \citenamefont {Roati}}]{burchianti2014efficient}%
  \BibitemOpen
  \bibfield  {author} {\bibinfo {author} {\bibfnamefont {A.}~\bibnamefont
  {Burchianti}}, \bibinfo {author} {\bibfnamefont {G.}~\bibnamefont
  {Valtolina}}, \bibinfo {author} {\bibfnamefont {J.}~\bibnamefont {Seman}},
  \bibinfo {author} {\bibfnamefont {E.}~\bibnamefont {Pace}}, \bibinfo {author}
  {\bibfnamefont {M.}~\bibnamefont {De~Pas}}, \bibinfo {author} {\bibfnamefont
  {M.}~\bibnamefont {Inguscio}}, \bibinfo {author} {\bibfnamefont
  {M.}~\bibnamefont {Zaccanti}},\ and\ \bibinfo {author} {\bibfnamefont
  {G.}~\bibnamefont {Roati}},\ }\href@noop {} {\bibfield  {journal} {\bibinfo
  {journal} {Physical Review A}\ }\textbf {\bibinfo {volume} {90}},\ \bibinfo
  {pages} {043408} (\bibinfo {year} {2014})}\BibitemShut {NoStop}%
\bibitem [{\citenamefont {Deshmukh}(2023)}]{Deshmukh2023}%
  \BibitemOpen
  \bibfield  {author} {\bibinfo {author} {\bibfnamefont {A.}~\bibnamefont
  {Deshmukh}},\ }\href@noop {} {\bibinfo {title} {A model of quantum
  diffractive heating}},\ \bibinfo {howpublished} {private communication}
  (\bibinfo {year} {2023}),\ \bibinfo {note} {talk at the Quantum Sensors for
  Vacuum Metrology, Vancouver, Canada, May 12, 2023.}\BibitemShut {Stop}%
\bibitem [{\citenamefont {Barker}\ \emph {et~al.}(2022)\citenamefont {Barker},
  \citenamefont {Acharya}, \citenamefont {Fedchak}, \citenamefont {Klimov},
  \citenamefont {Norrgard}, \citenamefont {Scherschligt}, \citenamefont
  {Tiesinga},\ and\ \citenamefont {Eckel}}]{barker2022precise}%
  \BibitemOpen
  \bibfield  {author} {\bibinfo {author} {\bibfnamefont {D.~S.}\ \bibnamefont
  {Barker}}, \bibinfo {author} {\bibfnamefont {B.~P.}\ \bibnamefont {Acharya}},
  \bibinfo {author} {\bibfnamefont {J.~A.}\ \bibnamefont {Fedchak}}, \bibinfo
  {author} {\bibfnamefont {N.~N.}\ \bibnamefont {Klimov}}, \bibinfo {author}
  {\bibfnamefont {E.~B.}\ \bibnamefont {Norrgard}}, \bibinfo {author}
  {\bibfnamefont {J.}~\bibnamefont {Scherschligt}}, \bibinfo {author}
  {\bibfnamefont {E.}~\bibnamefont {Tiesinga}},\ and\ \bibinfo {author}
  {\bibfnamefont {S.~P.}\ \bibnamefont {Eckel}},\ }\href@noop {} {\bibfield
  {journal} {\bibinfo  {journal} {Review of Scientific Instruments}\ }\textbf
  {\bibinfo {volume} {93}} (\bibinfo {year} {2022})}\BibitemShut {NoStop}%
\bibitem [{\citenamefont {Mitroy}\ and\ \citenamefont
  {Zhang}(2007)}]{mitroy2007}%
  \BibitemOpen
  \bibfield  {author} {\bibinfo {author} {\bibfnamefont {J.}~\bibnamefont
  {Mitroy}}\ and\ \bibinfo {author} {\bibfnamefont {J.-Y.}\ \bibnamefont
  {Zhang}},\ }\href {https://doi.org/10.1103/PhysRevA.76.032706} {\bibfield
  {journal} {\bibinfo  {journal} {Phys. Rev. A}\ }\textbf {\bibinfo {volume}
  {76}},\ \bibinfo {pages} {032706} (\bibinfo {year} {2007})}\BibitemShut
  {NoStop}%
\bibitem [{\citenamefont {Derevianko}\ \emph {et~al.}(2010)\citenamefont
  {Derevianko}, \citenamefont {Porsev},\ and\ \citenamefont
  {Babb}}]{derevianko2010}%
  \BibitemOpen
  \bibfield  {author} {\bibinfo {author} {\bibfnamefont {A.}~\bibnamefont
  {Derevianko}}, \bibinfo {author} {\bibfnamefont {S.~G.}\ \bibnamefont
  {Porsev}},\ and\ \bibinfo {author} {\bibfnamefont {J.~F.}\ \bibnamefont
  {Babb}},\ }\href@noop {} {\bibfield  {journal} {\bibinfo  {journal} {Atomic
  Data and Nuclear Data Tables}\ }\textbf {\bibinfo {volume} {96}},\ \bibinfo
  {pages} {323} (\bibinfo {year} {2010})}\BibitemShut {NoStop}%
\bibitem [{\citenamefont {Child}(1974)}]{child1974}%
  \BibitemOpen
  \bibfield  {author} {\bibinfo {author} {\bibfnamefont {M.~S.}\ \bibnamefont
  {Child}},\ }\href@noop {} {\emph {\bibinfo {title} {Molecular Collison
  Theory}}}\ (\bibinfo  {publisher} {{Academic Press, London and New York}},\
  \bibinfo {year} {1974})\BibitemShut {NoStop}%
\end{thebibliography}%

\end{document}